\documentclass[aps,pre,twocolumn,superscriptaddress]{revtex4-2}
\usepackage{graphicx}
\usepackage{dcolumn}
\usepackage{bm}
\usepackage{epsfig}
\usepackage{color}

\begin{document}


\title{Digging through two-dimensional densely-packed coarse granular
  media as a critical phenomenon}


  
\author{Guo-Jie Jason Gao}
\email{gao@shizuoka.ac.jp}
\affiliation{Department of Mathematical and Systems Engineering,
  Shizuoka University, Hamamatsu, Shizuoka 432-8561, Japan}

\author{Fu-Ling Yang} \affiliation{Department of Mechanical
  Engineering, National Taiwan University, Taipei 10617, Taiwan}

\author{Michael C. Holcomb} \affiliation{Department of Physics and
  Geosciences, Angelo State University, San Angelo, TX 76909-0904,
  USA}

\author{Jerzy Blawzdziewicz} \affiliation{Department of Physics, Texas
  Tech University, Lubbock, TX 79409-1051, USA}
\affiliation{Department of Mechanical Engineering, Texas Tech
  University, Lubbock, TX 79409-1021, USA}

\date{\today}

\begin{abstract}
We create a mechanical digger able to move within 2D densely-packed
granular media, in a manner intrinsically different from the existing
biomimic diggers. The characteristics of our design include that the
average grain size is about one tenth as large as the digger, and the
area packing density $\phi$ for testing is between about $0.610$ and
$0.762$. Unlike conventional autonomous diggers that move by
fluidizing the surrounding granular media with a much finer grain
size, our digger uses a different mobility mechanism as coarse grains
with interparticle friction are hard to be fluidized. To cope with
high interparticle friction, the digger has a circular shape and
singly captures granular particles near the front entrance of the
recess formed by the center unit running across its body and then
ejects the captured particles backwards. We validate this moving
strategy in both experiments using a manual digger and the
corresponding numerical simulations, and demonstrate that the moving
efficiency can be enhanced by the judgement of the human operator but
the effectiveness is not significantly influenced by the shape of
granular particles. In addition, localized vibration is needed to
degrade friction between interlocked irregular-shaped particles. Our
numerical results show that the distribution of both moving distances
and time intervals between consecutive ejections follow a
power-law. Reducing $\phi$ shrinks the spatial and temporal spans over
which the power laws hold. Further, we experimentally verify this
finding by demonstrating that similar power-laws are observed from an
automated digger which periodically randomizes its digging
direction. Finding these spatio-temporal power laws indicates that
digging within a densely-packed granular environment could be a
critical phenomenon.
\end{abstract}


\maketitle

\section{Introduction}
\label{introduction}

Digging and moving granular materials are common engineering processes
and have important industrial applications, but the facility design is
far from well-established due to the unique nature of the materials.
Systems composed of granular materials such as sand, gravel, and drift
ice are athermal and non-equilibrium because the energy required to
lift a single granular particle to a height comparable to its size
under gravity is usually much larger than the equilibrium thermal
fluctuation energy.  For thermal materials of much smaller
constituents, such as air and water, their equilibrium characteristics
permit us to treat them as a continuum and allow a universal governing
equation (e.g. the well-known Navier-Stokes equation) for predicting
the behavior of man-made objects moving within them.  Granular systems,
on the other hand, lack this kind of fundamental governing equation of
motion and usually stay still without a continuous injection of
external energy.  Granular materials can also transit from being
fluid-like to a disordered solid-like state if the packing density
becomes higher than the jamming density $\phi_{c}$ \cite{behringer19}
(for frictionless materials, $\phi_{c} \approx 0.84$ in
two-dimensional; for frictional materials, $\phi_{c}$ exists over a
broad, protocol-dependent range of packing densities
\cite{behringer15}). These bi-phasic complex features make designing a
digging machine (digger) mobile in dense athermal materials both
challenging and less explored.

Yet, novel manipulations of such material often lead to technological
advancement, and there have been a finite number of attempts that
challenge the design of a mobile digger for densely-packed granular
media. Except for a few artificial designs \cite{gao19,donado24}, the
majority is bio-inspired and mimics the movements of organisms
\cite{goldman14, goldman15, goldman16, astley20, fischer21,
  daltorio23} such as bacteria \cite{zhang12,zenit19}, snakes
\cite{goldman15_2, goldman19}, sandfish lizards \cite{herrmann09,
  goldman09, goldman13, goldman15_1, pak16}, bristled worms
\cite{tolley19}, earthworms \cite{anselmucci21, goldman22}, plant
roots \cite{hawkes18, naclerio21}, self-burying seeds \cite{tao24},
razor clams \cite{hosoi11, winter12, hosoi14, tao20, tao20_1}, or mole
crabs \cite{stuart22}. Generally, diggers in granular media can be
divided into two categories based on the dimensionless size $L/d$, an
important parameter affecting the mobility of the digger
\cite{goldman15, andrade24}, where $L$ is the digger's characteristic
size and $d$ is the characteristic grain size. All known man-made
autonomous diggers fall into the category of $L/d \gg 1$, where a
large number of grains are moved at once and react more like a
continuum for the digger. These bio-inspired diggers alternate their
shapes in prescribed manners, generating localized mobility by
disturbing the surrounding granular media from a jamming state to a
yielding one. However, this kind of strategy does not work in the
category of $L/d \approx \mathcal{O}(10)$, where only a few grains can
be moved at a time because the power of the digger is not enough to
fluidize/plasticize the granular medium in nearly jammed regions.
Specifically, when the size of a single grain is comparable to that of
the space created by the digger using its moving parts, which often
scales with $L$ or smaller, the interparticle friction becomes
difficult to overcome.  As a result, the surrounding granular
particles become a major obstacle that stops the digger from moving.

To address this problem, we design a digger that locally unjams
granular particles by reducing interparticle friction using small yet
focused vibrations and relocates coarse granular particles one at a
time by a repetitive procedure. Local friction reduction introduces
additional degrees of freedom as $\phi_{c}$ for a frictionless system
is greater than a frictional system
\cite{bi11,papanikolaou13,behringer15} so that the frictional particle
configuration in the vibrated neighborhood now appeared as in an
unjammed state. By locally unjamming particles, the digger not only
creates space but may also change the interparticle force network in
front of it to wiggle and push to displace particles in its way. The
newly-introduced vibration makes the current design an actively-driven
probe whose propulsion arises from both internal actuation and
vibration, which is different from conventional intruders passively
driven at constant velocity or force
\cite{behringer05,kozlowski19,franklin22,franklin26}.

This work exploits sequential experiments and simulations to test the
proposed digging strategy, validates the unjamming mechanism and pins
down the underlying physics. The proposed digger framework is
initially tested via manual operation. The manual digger implements
two mechanisms to capture a particle: 1) active selection by the human
operator and 2) wiggling and pushing followed by system-wide
relaxation. The propensity of the human operator to select particles
that are easy to capture prevents objective evaluation of the
effectiveness of the proposed mechanism. To tackle this issue, we
perform a discrete element simulation where the digger
indiscriminately captures particles by following a prescribed
procedure, containing small random displacements of the digger
followed by energy minimization of the whole system, removing the
influence of subjective decision-making. The reason behind imposing
random displacements for the digger is a granular system may jam if
loaded along a certain direction but unjam if pushed in a different
direction \cite{claudin98}. Our numerical results show that both the
moving distances and the time intervals of the digger between
consecutive captures and ejections of granular particles exhibit a
power-law distribution. To verify this finding by experiment with
minimal human influence, we further introduce a control protocol that
periodically assigns random digging directions within a fixed angular
range on the front side of the digger. Surprisingly, we find similar
spatio-temporal power laws with exponents around one. The numerical
and experimental power laws indicate that digging within a
densely-packed granular environment via random localized unjamming may
be a critical phenomenon \cite{boer95,ito95,drocco05}.

Below we introduce the design of our digger in section
\ref{experimental}, and show our experimental and numerical results of
operating the digger in 2D densely-packed granular media in section
\ref{results}. We conclude our study in section \ref{conclusions}.

\section{The Digger and Experimental Setup}
\label{experimental}

\begin{figure}
\includegraphics[width=0.39\textwidth]{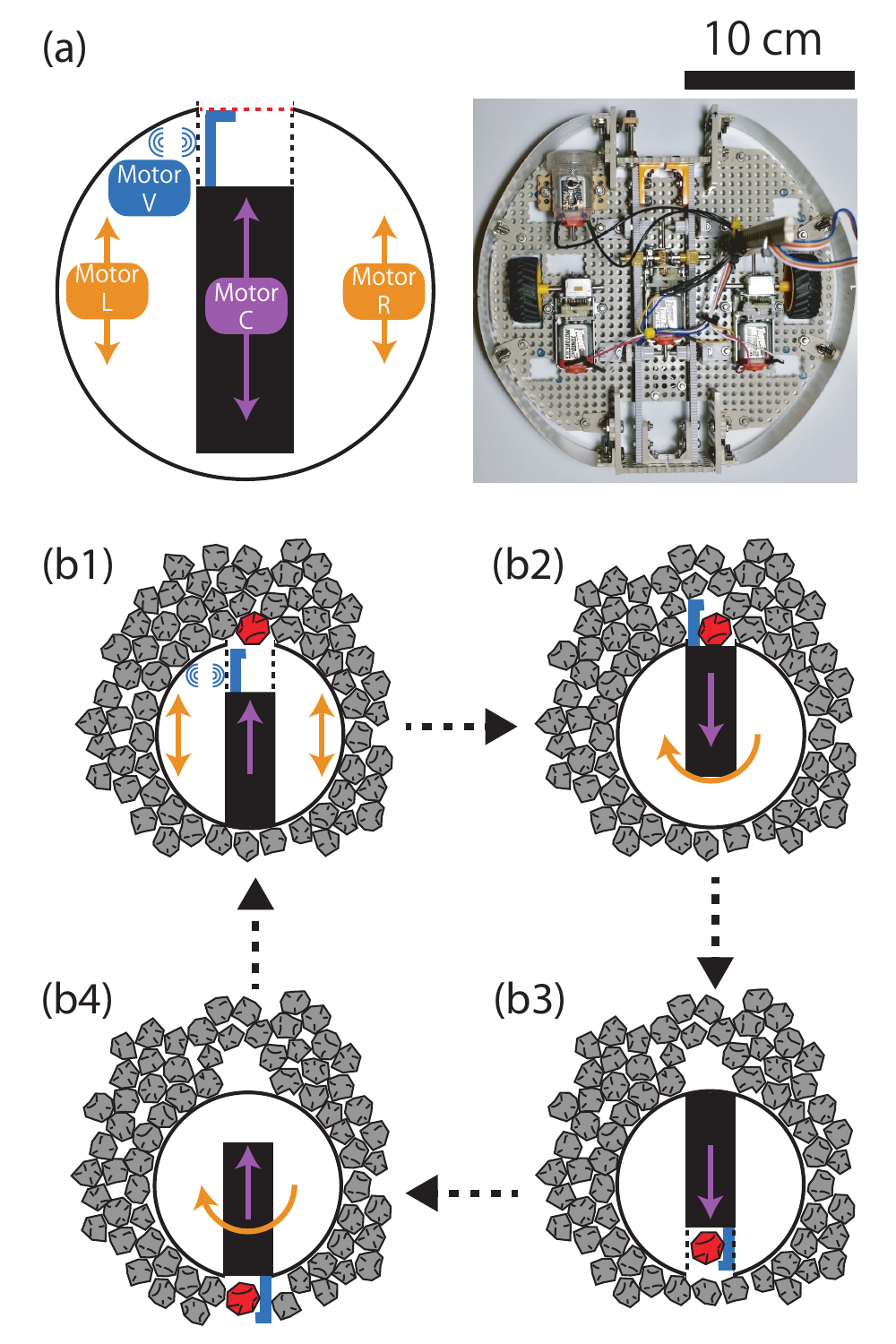}
\caption{\label{fig:schematic} (a) Schematic (left) and image (right)
  of the digger. Each driving wheel is independently powered by a
  motor, $L$ or $R$ (orange). The center unit equipped with an
  optional (if capturing irregular particles) claw is driven by a
  third motor $C$ (purple) while a fourth motor $V$ (blue) generates
  vibrations. (b) The repetitive four-step procedure of the digger
  moving through granular particles: 1. The digger captures the
  targeted particle (red) by wiggling and pushing using the two
  independent driving wheels (orange arrows) and/or using the
  vibrating function and the claw (blue) attached to the center unit
  (purple arrow). 2.  The digger then retracts the center unit,
  securing the particle inside of its recess, as it rotates by $180$
  degrees with a random rotation direction. 3. The digger ejects the
  captured particle using the center unit, completing the particle's
  relocation. 4. Finally, the digger resets its position and the
  procedure starts over.}
\end{figure}

We test our design experimentally by building a flat digger with a
circular shape equipped with an internal recess able to facilitate
localized vibrations, as shown schematically with an image in
Fig. \ref{fig:schematic}(a). More construction details of this digger
is provided in Sec. \ref{mech_details}. The circular shape allows the
digger to rotate and change its moving direction easily with minimal
interference with the adjacent granular particles. When the digger
moves, it creates small local displacements of multiple adjacent
particles. The digger can directly capture a particle into the
internal recess on its front edge or indirectly receives one through
the accumulation of small cooperative displacements from adjacent
particles. In other words, the internal recess allows the digger to
proceed without excavating a sizable tunnel into the bed - meaning
that the yielding of the surrounding granular particles is not a
concern, which is important when the packing density $\phi$ is close
to the jamming density $\phi_{c}$, an environment that is hard to
fluidize/plasticize.

\subsection{Moving procedure of the digger}
\label{moving_procedure}

The digger moves through a bed of granular particles by following an
iterative four-step procedure, as illustrated in
Fig. \ref{fig:schematic}(b). Upon activating the wheels' forward
motion, the digger wiggles and pushes to capture the granular
obstacles in front of it into the recess in the center unit
(Fig. \ref{fig:schematic}(b1)). Capturing irregularly-shaped granular
particles that interlock easily through this pushing and wiggling
process is impractical. To overcome this, the digger can use localized
vibrations to loosen the interlocked granular particles and then
capture them. To minimize human influence and verify the results from
numerical simulations, we automate this step by periodically assigning
random digging directions within a fixed angular range on the front
side of the digger, described in
Sec. \ref{automated_digger}. Additionally, after the digger has
wiggled and retracted its claw (if used) to capture a particle within
its recess, it rotates $180$ degrees in a random direction
(Fig. \ref{fig:schematic}(b2)).  A particle is treated as captured by
the digger if its geometric center is within the bounds of the
digger's recess, as indicated by the red dashed line shown in
Fig. \ref{fig:schematic}(a). In the experimental setup, the operator
of the automated digger makes this determination. After finishing its
rotation, the digger uses its center unit to eject the captured
particles backwards (Fig. \ref{fig:schematic}(b3)). Finally, the
center unit returns to its original configuration and then the digger
rotates by $180$ degrees to begin the moving procedure again
(Fig. \ref{fig:schematic}(b4)). With each iteration of the moving
procedure, the digger advances into the space left by the displaced
particles. This distance is much shorter than its size, but with
numerous repetitions of the procedure the digger can move a
considerable distance overall.

\subsection{Mechanical details of the digger}
\label{mech_details}

The circular digger measures $21\ \mathrm{cm}$ ($D_d$) in diameter, is
driven by two independent wheels, sitting symmetrically aside the
centerline of the digger. The two driving wheels allow the digger to
move forward, move backward, or pivot on a surface. Each wheel is
driven by a separate gearbox. Another center unit is installed along
the line of symmetry of the digger and driven by its own gearbox to
move linearly back and forth along a track. The center unit creates
the internal recess within and throughout the digger via an opening on
the front edge. All of the gearboxes are identical, equipped with an
FA-130 motor powered by two 1.2 V, 2,450 mAh Ni-MH batteries, and set
at a low-speed gear ratio of 719:1 to output a high torque. Another
FA-130 motor placed near the front of the digger generates localized
vibrations at a frequency of $f \approx 200$ Hz, which can be
transmitted to the granular obstacles by the center unit. Practically,
the motor giving vibrations is installed on the left side of the
digger, so strictly speaking, the digger is asymmetric. However, this
asymmetry has no qualitatively detectable impact on the digger’s
overall advancement function.  For better efficiency within
irregular-shaped granular particles, we attach an optional claw to the
center unit capable of delivering vibrations to a desired location. To
minimize any possible performance issues associated with low battery
voltage, we ensure that all batteries are drained by no more than
$10\%$ of the fully charged voltage of $1.2$ V for each
experiment. All parts are commercially available, produced by a
plastic model manufacturer, Tamiya.

\subsection{Vibration ability of the digger}
\label{digger_vib_ability}

Our digger can exert vibrations near the front edge of the central
recess unit at a frequency $f \approx 200$ hertz and an amplitude $A
\approx 0.042 \pm 0.001$ millimeters, evaluated by the peak-peak
amplitude of a vibrometer (SHOWA SOKKI, Model 1332B-00F and 2302B)
placed at the recess of the digger. In particular, the amplitude was
measured in the digger's moving direction every five seconds for one
minute. It converts to a dimensionless peak acceleration $\Gamma = A(2
\pi f)^2/g \approx 4.33 \pm 0.32$, where $g$ is the gravitational
acceleration. With vibrations, the interparticle friction is
significantly reduced around the recess. This allows the digger to
individually loosen, capture, and relocate interlocked gravel grains
blocking its way. Experimentally, using mechanical vibrations of lower
frequency ($50 - 300$ Hz) \cite{vidal15, melo17} or ultrasonic
vibrations ($20$ kHz) \cite{harkness17} with small amplitude
($\lesssim 10$ $\mu$m) has been reported to effectively reduce the
force resisting the slip of a slide when being pulled on a granular
layer or the penetration of an intruder into a dry granular medium
composed of grains much smaller than the intruder, the latter is an
effect known as acoustic fluidization \cite{melosh96,
  melo12}. However, there is as of yet no bio-inspired digger equipped
with a vibrating function at a frequency of a few hundred hertz since,
to the best knowledge of the authors, animals do not use vibrations at
this frequency to move forward in a packed granular medium.  The
vibrating function serves as a key component in creating the mobility
of our digger.  Our prototype digger would be the first to utilize
localized high-frequency vibrations in digging forward.

\subsection{Specifications of the experimental setup}
\label{setup_exp}

\begin{figure}
\includegraphics[width=0.45\textwidth]{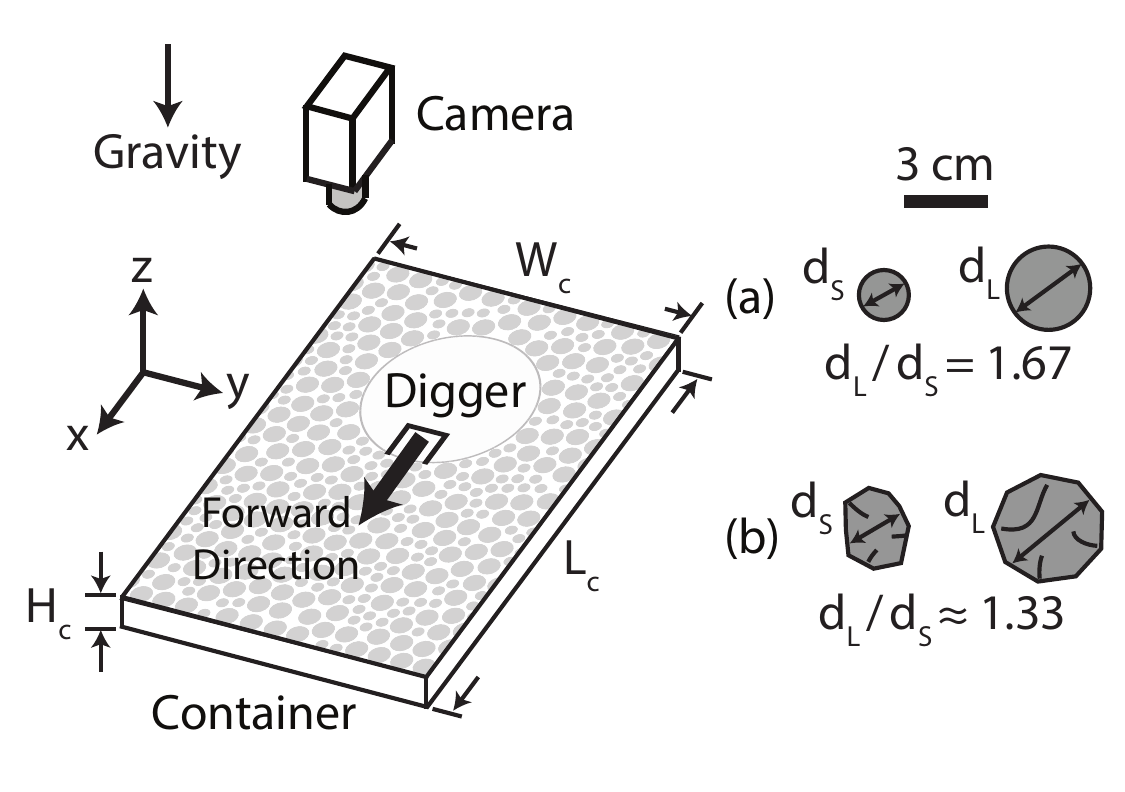}
\caption{\label{fig:setup} Schematic of the experimental setup.  The
  circular digger moves within a layer of densely-packed granular
  particles of area packing density close to the jamming packing
  density $\phi_{c}$. The digger and packed granular particles are
  confined in a wood container with an aluminum base measuring
  $37\ \mathrm{cm}$ in width ($W_c$), $54\ \mathrm{cm}$ in length
  ($L_c$), and $3\ \mathrm{cm}$ in height ($H_c$).  A digital camera
  records the motion of the digger on the $x-y$ plane from the top.
  For granular particles, we use (a) simple-shaped wood cylinders with
  a size ratio of $1.67$, and (b) irregular-shaped granite gravels
  with size ratio of $\approx 1.33$ to avoid crystallization. The
  dimensionless size $L/d$ between the digger and an average grain
  remains $\approx \mathcal{O}(10)$. }
\end{figure}

We place the digger in a layer of densely-packed granular particles of
area packing density close to $\phi_{c}$ and $L/d \approx
\mathcal{O}(10)$, as shown schematically in Fig. \ref{fig:setup}. The
digger can only move along a flat surface and is not allowed to climb
over the bed of granular particles. In other words, the current test
environment contains no freely moving boundary surface and therefore
is harder for the digger to move around, unlike our previous work
\cite{gao19}. Specifically, we prepared a simple two-dimensional test
bed, a wood container with an aluminum base that measures
$37\ \mathrm{cm}$ ($W_c$) in width, $54\ \mathrm{cm}$ ($L_c$) in
length, and $3\ \mathrm{cm}$ ($H_c$) in height and accommodates a
layer of $N$ densely-packed granular particles and our digger. For
each experiment, we manually fill an empty container with the desired
amount of granular particles and shake the whole container for $10$
seconds at a frequency $f \approx 80$ Hz and a dimensionless
acceleration $\Gamma \approx 7.66 \pm 0.19$ using a vibrator with an
input power of $650\ \mathrm{W}$.  The positions $(x,y)$ of the digger
are captured over time by a digital camera from atop of the container
to evaluate the digger's mobility.

To test the effect of particle shape on the performance of the digger,
we use two kinds of bidisperse granular particles: simple-shaped wood
cylinders and irregular-shaped granite gravels, both satisfying an
$L/d \approx \mathcal{O}(10)$ with $L=D_d$ and $d=d_S$, the diameter
of small particles.  Since the digger exhibits only in-plane movement,
we use the 2D area packing density $\phi_{2D}$ to quantify the
compactness of granular particles.  For a system containing $N$
cylinders of diameter $d_i$ and base area $a_i=\pi {{({d_i}/2)}^2}$,
$\phi_{2D}$ is defined as
\begin{equation} \label{2D_phi}
{\phi _{2D}} = \frac{{\sum\limits_{i = 1}^N {{a_i}} }}{{[{W_c}{L_c} -
      {A_d}]}},
\end{equation}
where $A_d=\pi {{({D_d}/2)}^2}$ is the base area of the circular
digger.  The cylinders are 50-50 (by number) bidisperse with a
diameter of $d_S = 1.8\ \mathrm{cm}$ or $d_L = 3.0\ \mathrm{cm}$ and a
diameter ratio of $1.67$ to avoid crystallization. When using
cylindrical particles, we choose $\phi_{2D} \approx 0.610$, $0.686$,
and $0.762$ to investigate the effect of $\phi_{2D}$ while making sure
that the system is dominated by the interparticle friction and
difficult to fluidize. The initial configurations with $\phi_{2D}
\approx 0.610$ and $0.686$ are prepared by first preparing
configurations with $\phi_{2D} \approx 0.762$ and then randomly
removing $20\%$ and $10\%$ particles, respectively, so that a lower
$\phi_{2D}$ can be generated with minimal structural differences.

For a system containing irregular-shaped granite gravels where
$\phi_{2D}$ is not well defined, we treat the system as a quasi-2D
porous medium to estimate a characteristic total area $a^{tot}$ for
all gravels by $a^{tot} = {w^{tot}}/(\rho_p \bar h)$
\cite{jung10}. Here, $w^{tot} = \sum_{i = 1}^{N} {w_i}$ is the total
weight of all granite gravels; $\rho_p = 2,500\ \mathrm{kg/m^3}$ is
the density of granite; and $\bar h$ is the average height of the
quasi-2D gravel layer. To measure $\bar h$, we divided the container
into an equally-spaced 7-by-7 matrix, measured the height $h$ of each
segment, and took the average of all 49 $h$'s as a global average
$\bar h$ (or first took the average of every 7 $h$'s and then took the
average of the seven groups of $h$ as a local average $\bar h$). Once
$w^{tot}$ and $\bar h$ are measured, the packing density ${\tilde \phi
}_{2D}$ of the porous granular medium is estimated by
\begin{equation} \label{quasi2D_phi}
{{\tilde \phi }_{2D}} = \frac{{w^{tot}}/(\rho_p \bar h)}{{[{W_c}{L_c}
      - {A_d}]}}.
\end{equation}
Similar to the wood cylinders, the gravel is roughly 50-50 (by number)
bidisperse with an estimated mean diameter of $d_S = 3.0\ \mathrm{cm}$
or $d_L = 4.0\ \mathrm{cm}$ and diameter ratio of $\approx 1.33$ to
avoid potential crystallization or excessive local void.  We choose
${\tilde \phi }_{2D} \approx 0.783$ using the global average $\bar h$
(or ${\tilde \phi }_{2D} \approx 0.774 \pm 0.107$, if $\bar h$ is the
local average with variations) when testing with the irregular-shaped
particles. This value was chosen to be close to that tested with the
regular-shaped cylinders as $\phi_{c}$ for the irregular-shaped
particles is not well-defined.

\section{Results and Discussion}
\label{results}

We first test the manually-operated digger within a layer of
bidisperse regular or irregular shaped granular particles. Next, we
conduct corresponding discrete element simulations using a specific
advancement protocol, which is later implemented in the final
experiment to automate the digger to confirm objectivity of the
digging mechanism. For each case, we found dynamically similar results
for the digger’s mobility using ten realizations in simulation and
three realizations in experiment.

\subsection{The manual digger in a layer of cylindrical particles}
\label{cylindrical_part}

\begin{figure}
\includegraphics[width=0.39\textwidth]{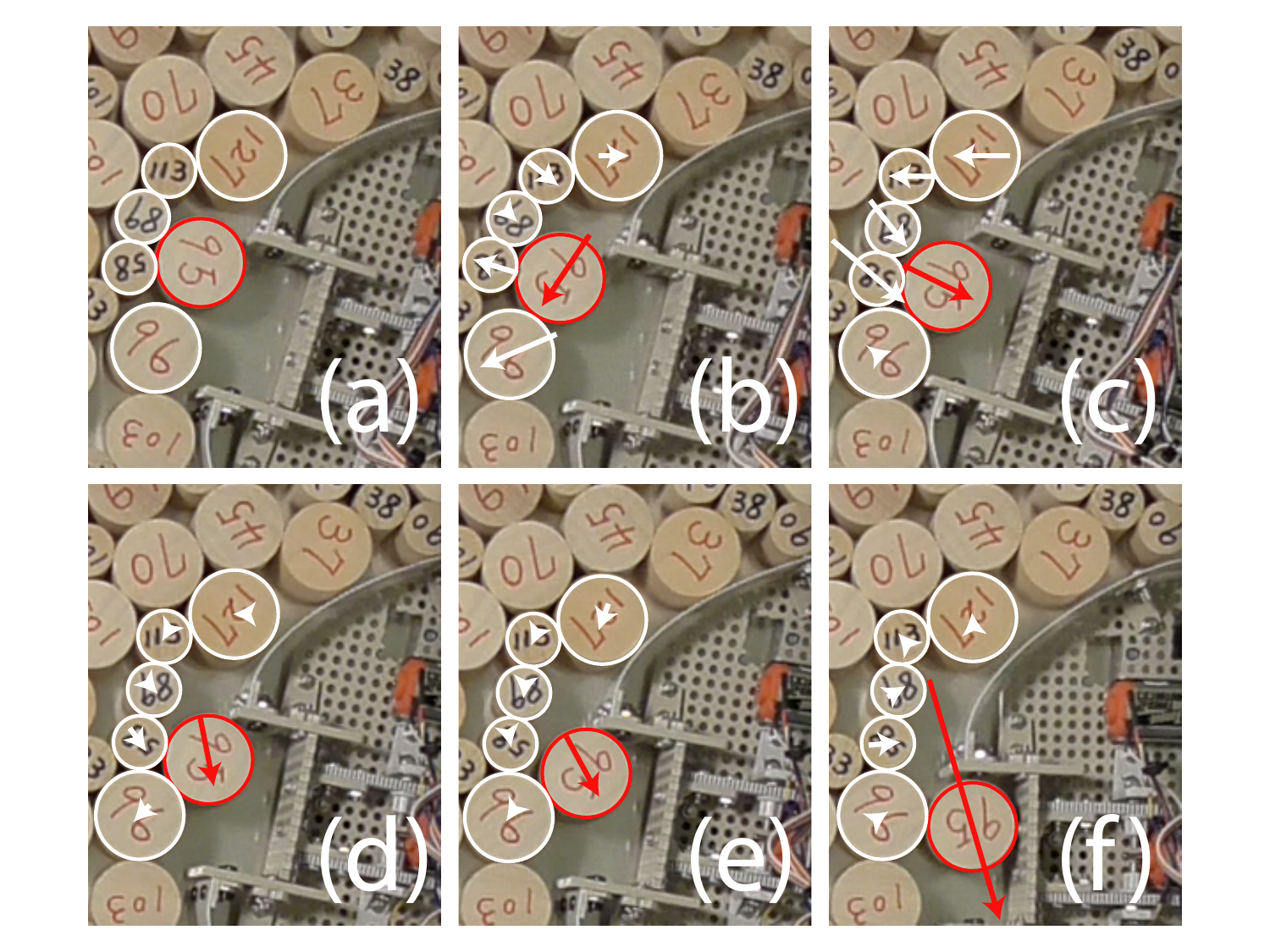}
\caption{\label{fig:Exp_cylinder_6th_L95_details} Time-lapse of the
  first and the second step of the repetitive moving procedure shown
  in Fig. \ref{fig:schematic}(b1) and (b2) with the system packed with
  $\phi_{2D} \approx 0.762$. (a-b) The digger wiggles and pushes to
  disturb the local network of a targeted particle (red-circled) and
  its neighbors (white-circled). (c) The digger pushes the neighbors
  of the targeted particle, whose collective responses cause it to
  move into the front recess of the digger. (d-f) The digger rotates
  to capture the targeted particle inside of its recess. Added arrows
  show each particle's direction of movement between frames, with
  length being proportional to the distance moved.}
\end{figure}

\begin{figure}
\includegraphics[width=0.34\textwidth]{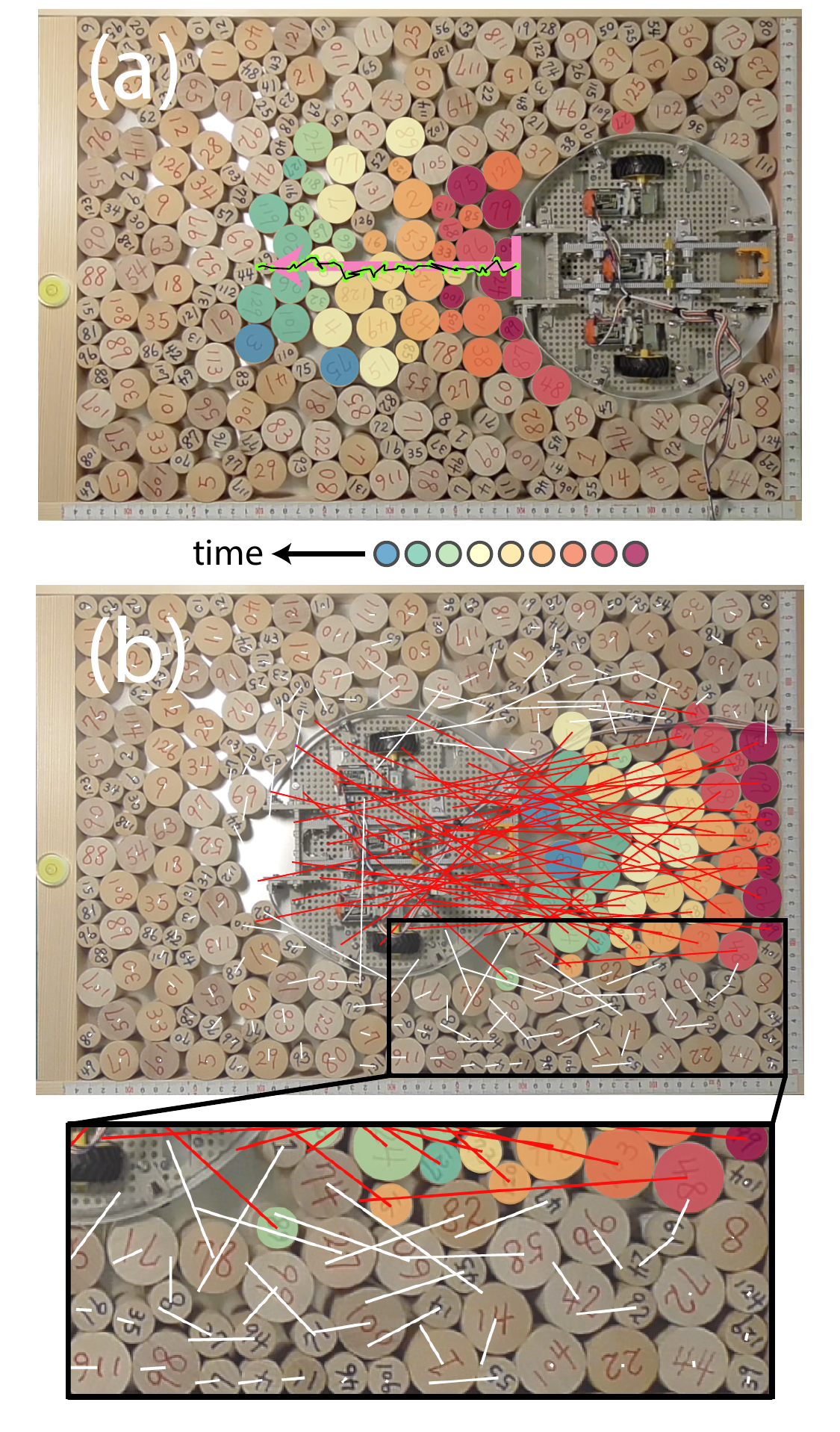}
\caption{\label{fig:Exp_cylinder_N50moved} Experimental packing
  configurations of the system (a) before and (b) after the manual
  digger has ejected a total of $50$ cylindrical granular particles
  and moved a distance (pink arrow) comparable to its size without
  using vibrations. The total elapsed time is $2790$ seconds. The
  actual trajectory of the digger after each ejection is also shown
  (green dots connected by black lines). System packed with $\phi_{2D}
  \approx 0.762$.  Relocated particles are colored in chronological
  order from red to blue and were subjectively selected by the
  operator of the digger.  The straight lines in (b) show the
  center-to-center displacement of each particle relocated by the
  digger (red), or merely disturbed by it (white), between the states
  (a) and (b). The blowup shows details of the lower-right corner,
  where particles closer to the boundary of the container are less
  disturbed.}
\end{figure}

The simplest experimental setup to validate the repetitive moving
procedure is to use circular granular particles.  We first test the
manually-operated digger within a layer of bidisperse wood cylinders
of $\phi_{2D} \approx 0.762$. With no chance of interlocking between
particles, the manual digger can proceed without using vibrations. An
explanatory time-lapse images showing how the digger captured and
secured a granular particle is shown in
Fig. \ref{fig:Exp_cylinder_6th_L95_details}, where the captured
particle is marked in red while the neighbors that show cooperative
motions are marked in white, with their displacements measured by
visual tracking. Its corresponding movie clip is provided in the
Supplemental Material \cite{supp1}, where the digger wiggles and
pushes to disturb the local network formed between the targeted
particle and its neighbors. The targeted particle is indirectly pushed
into the digger's front recess as its neighboring particles undergo
plastic reorganization. This process takes place as the accumulation
of small cooperative motions among the involved particles, which
collectively exhibit a fragile mechanical response
\cite{claudin98}. For the digger to move a distance comparable to its
size, the number of particles to be relocated is at the order of
$(D_d/d_L)^2$. The initial configuration of the system is shown in
Fig. \ref{fig:Exp_cylinder_N50moved}(a), where particles are colored
from red to blue according to the chronological order in which they
were relocated during the operation.
Fig. \ref{fig:Exp_cylinder_N50moved}(b) shows the final configuration
of the system after the movement of the digger, where the straight
lines show the displacement between the two configurations of each
particle, either relocated (red) or merely disturbed (white) by the
digger.

Particles close to the digger must be moved earlier than particles
farther away from the digger.  Additionally, only particles relocated
by the digger have long displacements comparable to, or slightly
longer than, the diameter of the digger.  Particles disturbed by the
digger during its motion mostly have very short displacements and stay
around their initial positions.  The few exceptions are the result of
the human operator judging that disturbing them would enhance
mobility.  In other words, global collective particle rearrangements
rarely happen in this localized rearrangement strategy.  We also
notice that the transverse width of the spatial distribution of
relocated particles, perpendicular to the average moving direction of
the digger, is similar to the diameter of the digger, which is
inherent in the digger's 180-degree rotation protocol in the moving
procedure.

\subsection{The manual digger in a layer of irregular-shaped particles}
\label{irregular_part}

\begin{figure}
\includegraphics[width=0.34\textwidth]{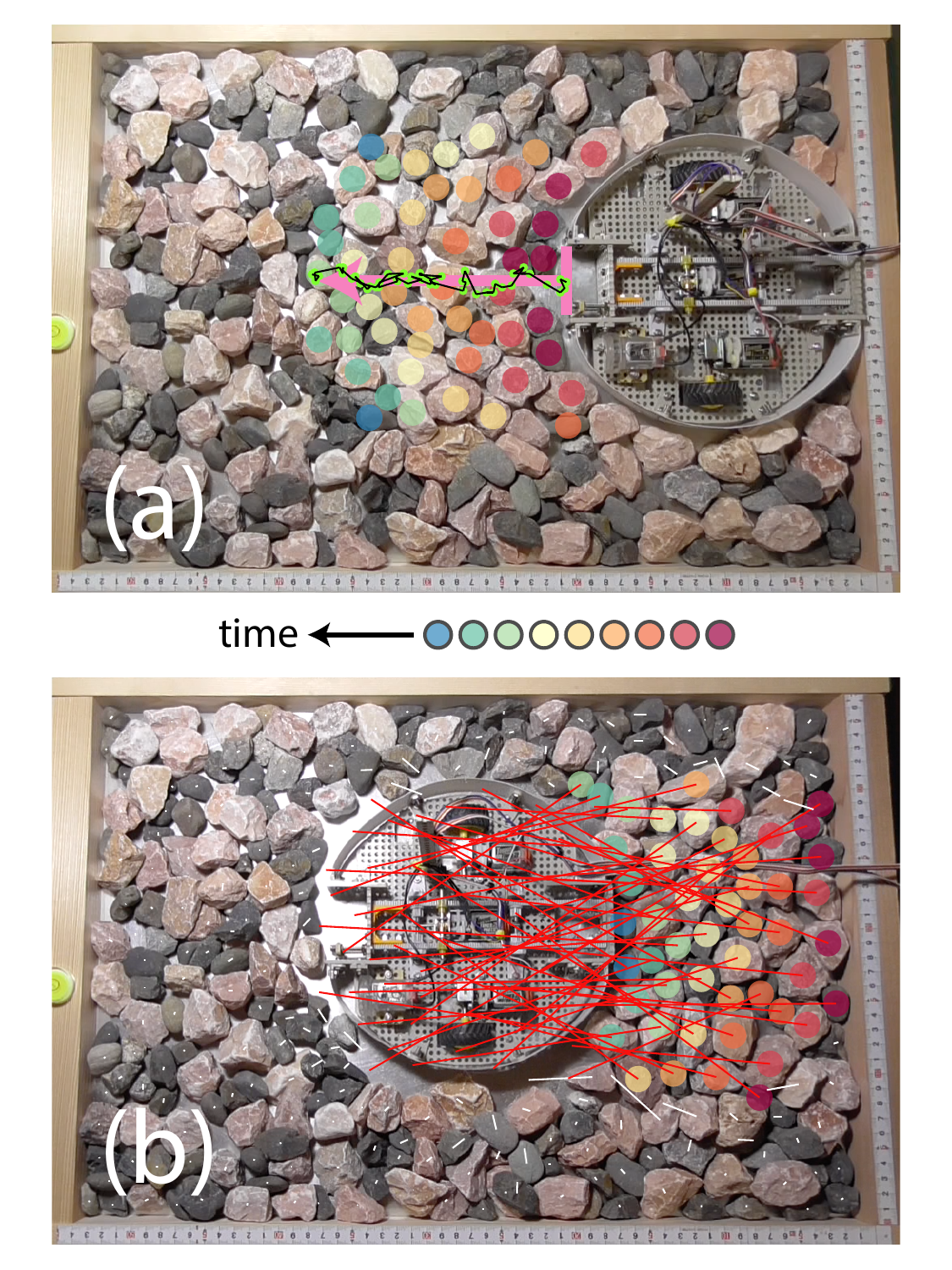}
\caption{\label{fig:Exp_gravel_N50moved} Experimental packing
  configurations of the system (a) before and (b) after the manual
  digger has moved a distance comparable to its size. The total
  elapsed time is $6177$ seconds.  The setup is similar to
  Fig. \ref{fig:Exp_cylinder_N50moved}, except particles are
  irregular-shaped, packed with ${\tilde \phi }_{2D} \approx 0.783$,
  and vibrations, delivered by the optional claw attached to the
  center unit, are used.}
\end{figure}

Having verified that the moving procedure does work for a layer of
bidisperse simple-shaped cylinders, we further tested the influence of
particle geometry on the digger's mobility by using irregular-shaped
bidisperse gravels of packing density ${\tilde \phi }_{2D} \approx
0.783$ (similar to that of the wood cylinders with $\phi_{2D} \approx
0.762$). We do find that vibrations are necessary for the digger to
move through irregular-shaped granular particles; a representative
movie clip is provided in the Supplemental Material \cite{supp2}. The
spatial distribution of relocated particles before and after the
digger has moved a distance comparable to its size are shown in
Fig. \ref{fig:Exp_gravel_N50moved}(a) and
Fig. \ref{fig:Exp_gravel_N50moved}(b), respectively.  Particles are
colored chronologically in the same fashion as in
Fig. \ref{fig:Exp_cylinder_N50moved}.  The motion of the digger within
irregular-shaped gravels is much slower, by about $2.2$ times ($6177$
seconds compared with 2790 seconds after relocating $50$ particles),
than within simple-shaped cylinders as there are interlockings between
gravels to be overcome. The results show that increasing the
complexity of particle shape notably hinders the mobility of the
digger. However, the spatial distribution of the particles relocated
before and after the digger's forward advancement is similar in
Fig. \ref{fig:Exp_cylinder_N50moved} and
Fig. \ref{fig:Exp_gravel_N50moved}. The proposed moving procedure is
indeed the key mechanism that permits the mobility of our digger in
the small $L/d$ category.

\subsection{The numerical digger using the discrete element method}
\label{numerical_digger}

\subsubsection{Details of the numerical digger}
\label{details_numerical_digger}

To provide more details of the digging process and quantitatively
analyze the digger's performance, we conducted two-dimensional
simulation using the discrete element method (DEM) in which the
digging process was designed to mimic the key procedures of the
mechanical digger. For each numerical simulation, we prepare a
bidisperse initial configuration by randomly placing points in the
container and then quasistatically growing them followed by energy
minimization of the whole system to reach the desired packing density;
the full details can be found in our previous study \cite{gao19}. In
the DEM simulation, the numerical digger pushes quasi-statically
forward successively until at least the center of one particle
naturally falls into the area of its front recess, as shown by the
movie clip provided in the Supplemental Material \cite{supp3}. This
ensures that the particle is captured without a subjective preference
by the operator, and our DEM model provides a means of objective
analysis. Then the numerical digger places the captured particle to
its rear, and ejects it by the modeled elastic normal force, without
the rotations of the digger in the two steps of (b2) and (b4) depicted
in Fig. \ref{fig:schematic}. The ejection force formula is defined in
Eqn. \ref{elastic_normal_force} below, which functions like the
mechanical center unit. The DEM simulation shall provide qualitative
responses to the experimental system. The details of the DEM algorithm
of the numerical digger are as follows.

First, with all the cylinders' positions fixed, the digger pushes
cylinders around it by changing its position from the current
$(x^{old},y^{old})$ to the new $(x^{new},y^{new})$ as follows:
\begin{equation} \label{digger_bump}
\begin{array}{l}
{x^{new}} = {x^{old}} + \Delta x\\ {y^{new}} = {y^{old}} + \Delta y
\end{array},
\end{equation}
where $\Delta x$ and $\Delta y$ are small random displacements
generated by two independent normal distribution functions $\left|
{{N_x}(0,{\alpha _x}\sigma )} \right|$ and ${N_y}(0,{\alpha _y}\sigma
)$ having zero means and standard deviations ${\alpha_x}\sigma$ and
${\alpha_y}\sigma$, respectively. The absolute value of $N_x$ makes
sure that the digger pushes forward only. We choose a small value
$\sigma = 0.005D_d$ so that the digger moves
quasi-statically. Throughout the study, we set $\alpha_x=5.0$ and
$\alpha_y=2.0$.

Second, the above quasi-static movement (push) of the digger is
followed by energy minimization (relaxation) of the whole system to
remove overlaps between objects, where Newton's translational
equations of motion are integrated using the velocity Verlet algorithm
\cite{tildesley17} until both the total potential energy and the total
kinetic energy of the system become sufficiently small. For
simplicity, Newton's rotational equations of motion are ignored
\cite{gao19}. The digger can move forward by repeating the two-step
procedure of quasi-static movement followed by energy
minimization. During the course of quasi-static movement, the digger
relocates granular particles whose centers fall into its recess area
to its rear by a distance of $D_d/2+3d_s$, where $d_s$ is the diameter
of small particles, which results in the ejection of captured
particles due to the modeled pairwise linear spring potential defined
in Eqn. \ref{elastic_normal_force}, which fulfills the task of the
mechanical center unit.

In the DEM simulation, each frictional granular particle $i$ obeys
Newton's translational equation of motion
\begin{equation} \label{newton_law}
{m_i}{{\mathord{\buildrel{\lower3pt\hbox{$\scriptscriptstyle\rightharpoonup$}}
      \over a} }_i} =
{{\mathord{\buildrel{\lower3pt\hbox{$\scriptscriptstyle\rightharpoonup$}}
      \over F} }_i} =
\mathord{\buildrel{\lower3pt\hbox{$\scriptscriptstyle\rightharpoonup$}}
  \over F} _i^N +
\mathord{\buildrel{\lower3pt\hbox{$\scriptscriptstyle\rightharpoonup$}}
  \over F} _i^C +
\mathord{\buildrel{\lower3pt\hbox{$\scriptscriptstyle\rightharpoonup$}}
  \over F} _i^D,
\end{equation}
where particle $i$ with mass $m_i$ accelerates by
${{\mathord{\buildrel{\lower3pt\hbox{$\scriptscriptstyle\rightharpoonup$}}
      \over a} }_i}$ due to the total force
${{\mathord{\buildrel{\lower3pt\hbox{$\scriptscriptstyle\rightharpoonup$}}
      \over F} }_i}$, composed of
$\mathord{\buildrel{\lower3pt\hbox{$\scriptscriptstyle\rightharpoonup$}}
  \over F} _i^N$,
$\mathord{\buildrel{\lower3pt\hbox{$\scriptscriptstyle\rightharpoonup$}}
  \over F} _i^C$, and
$\mathord{\buildrel{\lower3pt\hbox{$\scriptscriptstyle\rightharpoonup$}}
  \over F} _i^D$, which are forces acting on the particle from its
contact neighbors, the container, and the digger, respectively. Below
we give details of each of these constituent forces.

To simulate frictional granular materials, we consider the
interparticle inelastic damping forces in both normal and tangential
directions and the elastic repulsive force only in the normal
direction for simplicity. The interparticle force
$\mathord{\buildrel{\lower3pt\hbox{$\scriptscriptstyle\rightharpoonup$}}
  \over F} _i^N$ on particle $i$ having $n_c$ contact neighbors $j$
can be expressed as
\begin{equation} \label{interparticle_force_law}
\mathord{\buildrel{\lower3pt\hbox{$\scriptscriptstyle\rightharpoonup$}}
  \over F} _i^N = \sum\limits_{j \ne i}^{{n_c}}
        {[\mathord{\buildrel{\lower3pt\hbox{$\scriptscriptstyle\rightharpoonup$}}
              \over f} _{ij}^n} ({r_{ij}}) +
          \mathord{\buildrel{\lower3pt\hbox{$\scriptscriptstyle\rightharpoonup$}}
            \over f} _{ij}^{{d_n}}({r_{ij}}) +
          \mathord{\buildrel{\lower3pt\hbox{$\scriptscriptstyle\rightharpoonup$}}
            \over f} _{ij}^{{d_t}}({r_{ij}})],
\end{equation}
where
$\mathord{\buildrel{\lower3pt\hbox{$\scriptscriptstyle\rightharpoonup$}}
  \over f} _{ij}^n$ is the interparticle elastic normal force defined
in Eqn. \ref{elastic_normal_force}, and
$\mathord{\buildrel{\lower3pt\hbox{$\scriptscriptstyle\rightharpoonup$}}
  \over f} _{ij}^{{d_n}}$ and
$\mathord{\buildrel{\lower3pt\hbox{$\scriptscriptstyle\rightharpoonup$}}
  \over f} _{ij}^{{d_t}}$ are the interparticle inelastic normal and
tangential damping forces that deduct the kinetic energy of the
interacting particles after each pairwise collision, defined in
Eqn. \ref{particle_damping_force_n} and
Eqn. \ref{particle_damping_force_t}, respectively.

The interparticle elastic normal force between two particles $i$ and
$j$ along the unit vector ${{\hat n}_{ij}}$, pointing from the center
of particle $j$ to that of particle $i$, is governed by
\begin{equation} \label{elastic_normal_force}
\mathord{\buildrel{\lower3pt\hbox{$\scriptscriptstyle\rightharpoonup$}}
  \over f} _{ij}^n({r_{ij}}) = \frac{\epsilon }{{d_{ij}^2}}{\delta
  _{ij}}H ({\delta _{ij}}){{\hat n}_{ij}},
\end{equation}
where $r_{ij}$ is the center-to-center distance between particles $i$
and $j$, $\epsilon$ is the elastic force amplitude, $d_{ij} =
(d_i+d_j)/2$ is the average diameter of particles $i$ and $j$,
$\delta_{ij} = d_{ij} - r_{ij}$ is the interparticle overlap, and
$H(x)$ is the Heaviside step function. Also along ${{\hat n}_{ij}}$,
we consider the interparticle normal damping force proportional to the
relative velocity between particles $i$ and $j$
\begin{equation} \label{particle_damping_force_n}
\mathord{\buildrel{\lower3pt\hbox{$\scriptscriptstyle\rightharpoonup$}}
  \over f} _{ij}^{{d_n}}({r_{ij}}) = - {b_n}H ({\delta
  _{ij}})({{\mathord{\buildrel{\lower3pt\hbox{$\scriptscriptstyle\rightharpoonup$}}
      \over v} }_{ij}} \cdot {{\hat n}_{ij}}){{\hat n}_{ij}},
\end{equation}
where $b_n$ is the normal damping parameter and
${{\mathord{\buildrel{\lower3pt\hbox{$\scriptscriptstyle\rightharpoonup$}}
      \over v} }_{ij}}$ is the relative velocity between the two
particles. The interparticle tangential damping force has a similar
form
\begin{equation} \label{particle_damping_force_t}
\mathord{\buildrel{\lower3pt\hbox{$\scriptscriptstyle\rightharpoonup$}}
  \over f} _{ij}^{{d_t}}({r_{ij}}) = - {b_t}H ({\delta
  _{ij}})({{\mathord{\buildrel{\lower3pt\hbox{$\scriptscriptstyle\rightharpoonup$}}
      \over v} }_{ij}} \cdot {{\hat t}_{ij}}){{\hat t}_{ij}},
\end{equation}
where $b_t$ is the tangential damping parameter, and ${{\hat t}_{ij}}
= (-n_{ij}^y,n_{ij}^x)$ \cite{shattuck15}. We choose ${b_n} = {b_t} =
0.2\sqrt {{m_s}\epsilon } /{d_s}$ throughout our simulations
\cite{luding08}.

The interaction force
$\mathord{\buildrel{\lower3pt\hbox{$\scriptscriptstyle\rightharpoonup$}}
  \over F} _i^C$ between particle $i$ and the container includes
contributions from the side walls
$\mathord{\buildrel{\lower3pt\hbox{$\scriptscriptstyle\rightharpoonup$}}
  \over f} _i^W$ and the base plate
$\mathord{\buildrel{\lower3pt\hbox{$\scriptscriptstyle\rightharpoonup$}}
  \over f} _i^B$
\begin{equation} \label{part_box_force}
\mathord{\buildrel{\lower3pt\hbox{$\scriptscriptstyle\rightharpoonup$}}
  \over F} _i^C =
\mathord{\buildrel{\lower3pt\hbox{$\scriptscriptstyle\rightharpoonup$}}
  \over f} _i^W +
\mathord{\buildrel{\lower3pt\hbox{$\scriptscriptstyle\rightharpoonup$}}
  \over f} _i^B,
\end{equation}
where
$\mathord{\buildrel{\lower3pt\hbox{$\scriptscriptstyle\rightharpoonup$}}
  \over f} _i^W$ has an analogous form to the interparticle normal
interaction
$\mathord{\buildrel{\lower3pt\hbox{$\scriptscriptstyle\rightharpoonup$}}
  \over f} _{ij}^{{\mathop{\rm n}} }$, except with
$\epsilon^W=2\epsilon$, meaning when a particle hits a wall of the
container, it is equivalent to hitting its mirrored copy on the other
side of the wall. Additionally, the interparticle force
$\mathord{\buildrel{\lower3pt\hbox{$\scriptscriptstyle\rightharpoonup$}}
  \over f} _i^B$ is modeled in a self-propulsive manner
\begin{equation} \label{part_base_force}
\mathord{\buildrel{\lower3pt\hbox{$\scriptscriptstyle\rightharpoonup$}}
  \over f} _i^B = \mu ({v_0} - {v_i}){{\hat v}_i},
\end{equation}
where $\mu=5.0$ sets the magnitude of the self-propulsion force, $v_0$
is the target velocity of particle $i$, and $v_i$ is its current
velocity along its unit vector ${\hat v}_i$. We set $v_0=0$ to model
the friction from the base plate of the container, which eventually
stops the motion of particle $i$.

Lastly, the particle-digger interaction force
$\mathord{\buildrel{\lower3pt\hbox{$\scriptscriptstyle\rightharpoonup$}}
  \over F} _i^D$ takes the form of
$\mathord{\buildrel{\lower3pt\hbox{$\scriptscriptstyle\rightharpoonup$}}
  \over f} _{ij}^n$, where the index $j=D$, if a particle $i$ overlaps
with the edge of the digger. Otherwise, if a particle touches any
boundary of the front recess of the digger,
$\mathord{\buildrel{\lower3pt\hbox{$\scriptscriptstyle\rightharpoonup$}}
  \over F} _i^D$ takes the form of
$\mathord{\buildrel{\lower3pt\hbox{$\scriptscriptstyle\rightharpoonup$}}
  \over f} _i^W$.

Likewise, the digger is subject to the Newtonian reaction forces from
its contact granular particles and to self-propulsion force mimicking
the friction from the container's base plate, as defined in
Eqn. \ref{part_base_force}. The DEM simulations in this study use the
diameter $d_s$, the mass $m_s$ of the small particles, and the
interparticle elastic normal force amplitude $\epsilon$ as the
reference length, mass, and energy scales, respectively.

\subsubsection{The numerical digger in a layer of cylindrical particles within the container}
\label{DEM_cylindrical_part_wo_pbc}

\begin{figure}
\includegraphics[width=0.34\textwidth]{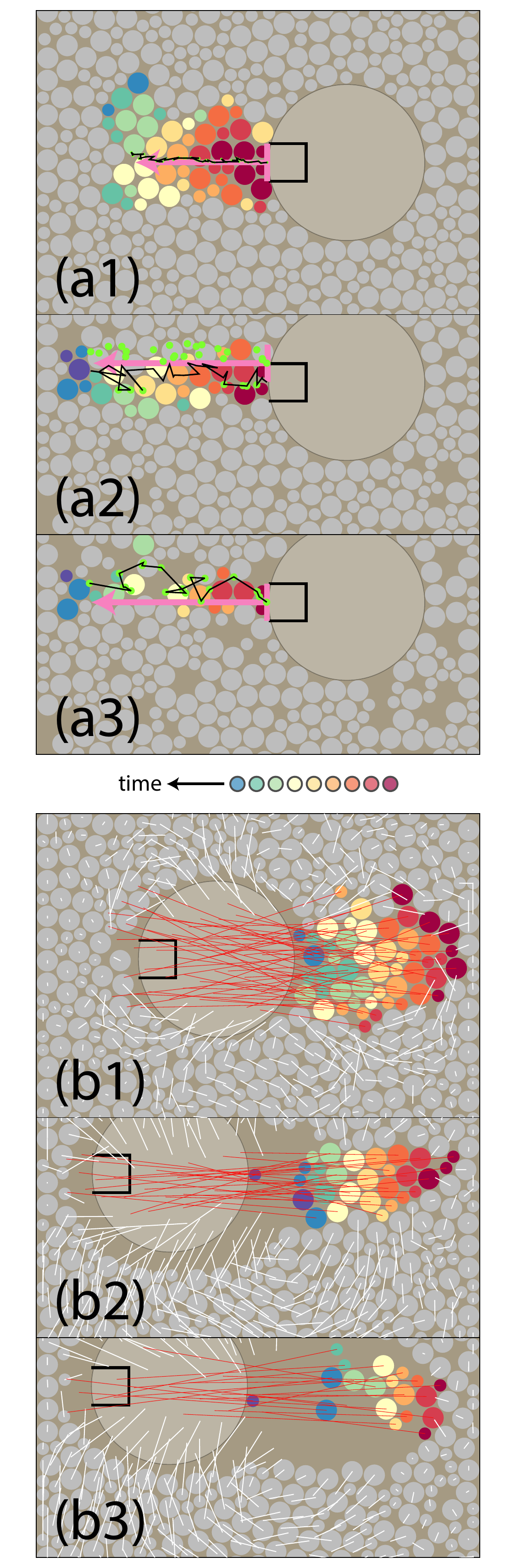}
\caption{\label{fig:Simu_N50moved} Packing configurations of the
  system (a) before and (b) after the numerical digger has moved a
  distance comparable to its size. System packed with $\phi_{2D}
  \approx 0.762$ (a1, b1), $0.686$ (a2, b2), and $0.610$ (a3, b3) and
  the total number of ejected particles $N_{eject}^{tot}$ are listed
  in Table \ref{tab:sim_exp_results}. The straight lines in (b) show
  the center-to-center displacement of each particle relocated by the
  digger (red), or merely disturbed by it (white), between the states
  (a) and (b).}
\end{figure}

The system configurations with $\phi_{2D} \approx 0.762$, $0.686$, and
$0.610$ before and after the digger has moved a distance comparable to
its size are shown in Fig. \ref{fig:Simu_N50moved}(a) and (b),
respectively.  The spatial distribution of relocated particles with
$\phi_{2D} \approx 0.762$ is strikingly similar to that found in the
manual experiment. In simulation, however, the relocated particles are
somewhat more confined in lateral width and the particle displacement
lines are distributed more symmetrically on the two lateral sides of
the numerical digger.  This occurs because the numerical digger moves
by the simple two-step procedure which always ejects particles along
its front-rear line of symmetry to its back.  In the manual
experiment, the human operator can selectively eject the captured
granular particles in any direction or merely choose to locally shift
them, as demonstrated by the particle displacement lines across the
digger in Fig. \ref{fig:Exp_cylinder_N50moved}(b). It may be argued
that human control demonstrates more flexibility because the operator
can control things like where to eject the captured cylinders to help
the digger adapt to the in-situ situation and move more efficiently.
Even though this kind of intentional adaptability cannot be easily
achieved within a numerical routine, the repetitive moving procedure
can still be confirmed effective for digger's mobility. Overall, a
lower $\phi_{2D}$ reduces the total number of ejected particles
$N_{eject}^{tot}$, listed in Table \ref{tab:sim_exp_results}, and
leaves bigger void behind the digger. In Sec. \ref{automated_digger}
and Fig. \ref{fig:Exp_cylinder_Microbit_N50moved}, we automate the
manual digger by periodically randomizing its digging direction, and
the spatial distributions of relocated particles are also very similar
to those found in the numerical experiments.

\begin{figure}
\includegraphics[width=0.39\textwidth]{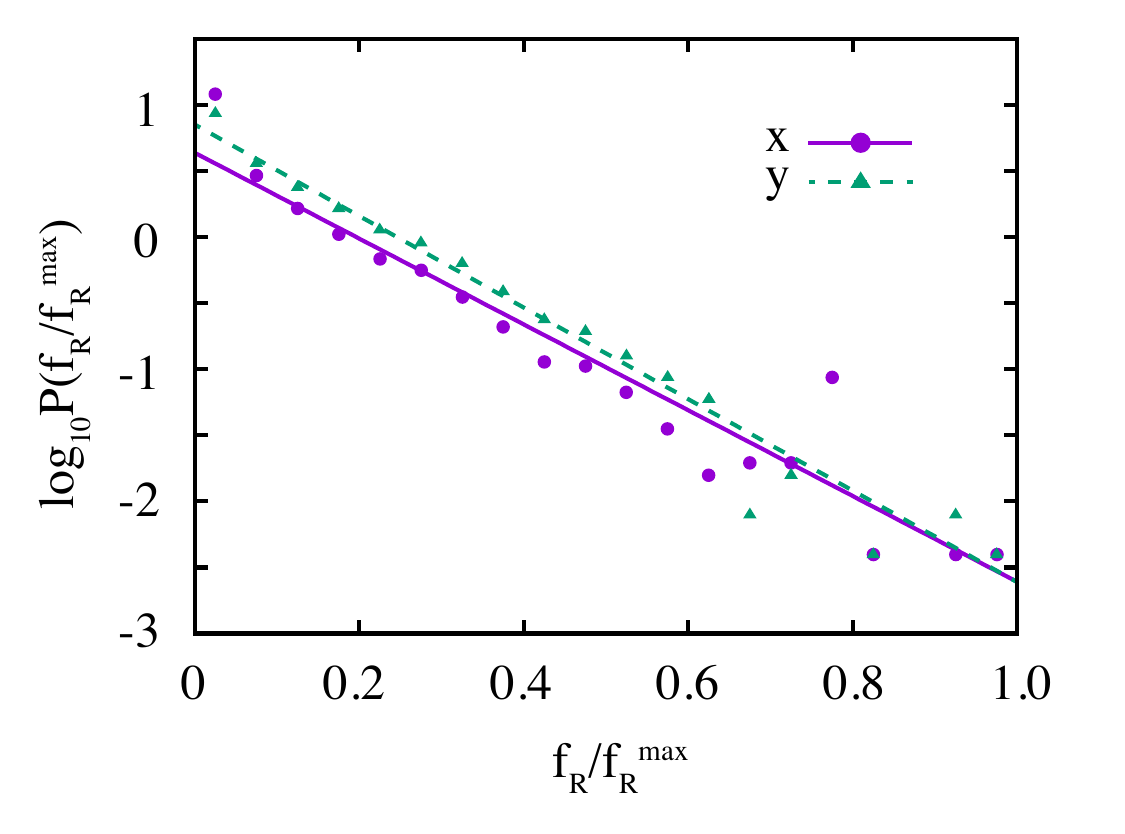}
\caption{\label{fig:Simu_i_fx_fy_info_pdf_analysis} The probability
  density distribution $P$ of resistance forces $f_R$, normalized by
  its maximal value $f_R^{max}$, experienced by the digger immediately
  after each quasi-static movement (push) and before the following
  energy minimization (relaxation) of the whole system after the
  digger has relocated about $(D_d/d_L)^2$ particles in the global
  reference direction ($x$, purple circles) and perpendicular to it
  ($y$, green triangles). The straight lines are linear regression
  with slopes -3.25 (purple) and -3.46 (green) of the distributed
  data. System packed with $\phi_{2D} \approx 0.762$.}
\end{figure}

To explore how difficult it is for the digger to move through
densely-packed granular obstacles, we examine the distribution of the
resistance force, $f_R = \sum_{i = 1}^{N} |
{\mathord{\buildrel{\lower3pt\hbox{$\scriptscriptstyle\rightharpoonup$}}
    \over F}_i^D} |$, experienced by the digger immediately after each
quasi-static movement and before the subsequent energy minimization of
the whole system for the case with a dense $\phi_{2D} \approx
0.762$. After the digger has relocated about $(D_d/d_L)^2$ particles,
the distribution of $f_R$, normalized by its maximal values
$f_R^{max}$, along the global reference direction $x$ or $y$ defined
in Fig. \ref{fig:setup} are shown in
Fig. \ref{fig:Simu_i_fx_fy_info_pdf_analysis}. We can see that the
distribution of $f_R/f_R^{max}$ spans several orders of magnitude on a
log-linear graph, and small forces develop way more frequently than
large forces. This differs from the distribution of contact forces in
granular media with a fall-off or plateau at very small forces
\cite{behringer05,majmudar05,richefeu09,voivret09}, but is extremely
similar to forces experienced by an actively growing plant root in a
granular packing, where passive interparticle force chains of a
granular medium translate into the resistance force experienced by an
active intruder \cite{fakih19}.  More importantly, the force
distribution along $x$, which is mostly aligned with the average
moving direction of the digger, exhibits a significant fluctuation in
the large force domain, indicating that occasionally the digger can
experience large resistance forces that make its motion uneven and
intermittent.

\begin{figure}
\includegraphics[width=0.39\textwidth]{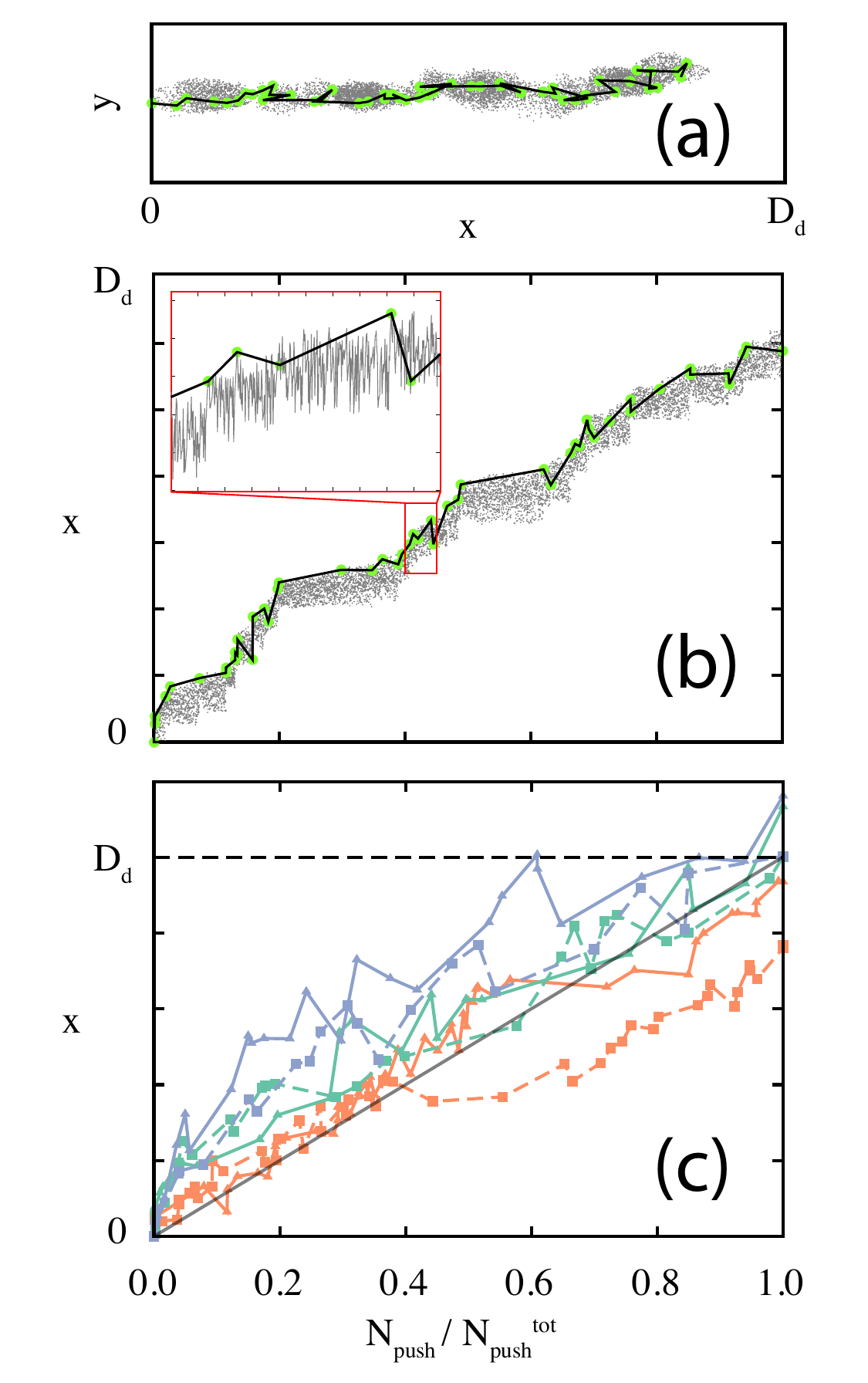}
\caption{\label{fig:Simu_N50moved_caging} (a) Digging dynamics of $50$
  ejections with $\phi_{2D} \approx 0.762$. Trajectory of the digger
  after each ejection (green dots connected by black lines) and the
  relaxed positions of the digger after each push (grey dots) are
  shown. (b) The $x$ components of the trajectory and the relaxed
  positions of the digger shown in (a) as a function of number of
  quasi-static pushes $N_{push}/N_{push}^{tot}$, normalized by
  $N_{push}^{tot}$ listed in Table \ref{tab:sim_exp_results} after the
  digger has moved a distance comparable to its size. The inset shows
  a magnified trajectory of the digger, where its discrete relaxed
  positions are linked by lines to indicate that the digger's being
  repeatedly pushed back by the neighboring particles. The stepwise
  shape of the curve reveals the digger's tendency to get stuck after
  ejecting several particles and provides clear indication of caging
  phenomenon. (c) Trajectories of the digger that are the nearest
  (squares connected by dashed lines) and the furthest (triangles
  connected by solid lines) from its initial position with $\phi_{2D}
  \approx 0.762$ (orange), $0.686$ (green), and $0.610$ (blue). The
  two trajectories for each $\phi_{2D}$ are selected from $10$
  realizations.}
\end{figure}

Simulation results also allow us to evaluate the digger's mobility in
terms of how far the digger has progressed and how many cylinders have
been relocated (via ejections) after the digger pushes forward
quasi-statically to attempt advancement, using the same numerical data
as in
Fig. \ref{fig:Simu_N50moved}. Fig. \ref{fig:Simu_N50moved_caging}(a)
and (b) examine the successful advancement trajectory (black lines)
with the relaxed positions of the digger (grey dots) after each
quasi-static push with $\phi_{2D} \approx 0.762$. In
Fig. \ref{fig:Simu_N50moved_caging}(a), We can distinguish several
blobs of relaxed positions, where the digger is confined locally for a
finite amount of time, demonstrating the caging effect
\cite{gao06}. This scenario recurs from time to time and is also
clearly reflected in the stepwise advancement of the digger along the
global reference direction $x$, when plotted against the total number
of pushes by the digger $N_{push}$ in
Fig. \ref{fig:Simu_N50moved_caging}(b). The simulation results show
that after ejecting several granular particles, the digger temporarily
gets stuck - that is, the digger ejects no particles in spite of many
advancement attempts where it quasi-statically pushes against the
neighboring particles. The blowup in
Fig. \ref{fig:Simu_N50moved_caging}(b) shows that the digger is
repetitively pushed back by particles blocking its way. The digger
subsequently becomes efficient again and can eject several more
particles before getting stuck once more. In
Fig. \ref{fig:Simu_N50moved_caging}(c), we compare the trajectories of
the digger along the $x$ direction against $N_{push}$ with $\phi_{2D}
\approx 0.762$, $0.686$, and $0.610$. The digger moves faster at the
initial stage with lower $\phi_{2D}$, and then it slows down while
approaching the wall of the container perpendicular to its moving
direction. The phenomenon is the most obvious with the lowest
$\phi_{2D} \approx 0.610$. In addition, the total number of
quasi-static pushes $N_{push}^{tot}$, listed in Table
\ref{tab:sim_exp_results}, decreases with $\phi_{2D}$, meaning a
faster digging.

\begin{table*}[t]
\caption{\label{tab:sim_exp_results} Variables relevant to the digger
  dynamics, averaged from $10$ numerical simulations (upper rows) or
  $3$ automated experimental realizations (lower rows) after the
  digger has moved a distance comparable to its size at different
  packing densities $\phi_{2D}$: $N_{eject}^{tot}$ is the total number
  of ejected particles, $N_{push}^{tot}$ is the total number of
  quasi-static pushes, $\Delta L^{max}$ and $\Delta N_{push}^{max}$
  are the maximum moving distance and maximum number of quasi-static
  pushes between two consecutive ejections, respectively, $m_{\Delta
    L}$ is the fitted slope on a log-log plot of $P(\Delta L / \Delta
  L^{max})$, and $m_{\Delta N}$ is the fitted slope on a log-log plot
  of $P(\Delta N_{push} / \Delta N_{push}^{max})$. In experiment,
  $\Delta N_{push}^{max} = \Delta T_{push}^{max}/(\Delta t_f + \Delta
  t_b)$, where $ \Delta t_f + \Delta t_b = 90 \text{ms}$. A
  single-valued $m_{\Delta L}$ or $m_{\Delta N}$ in the parenthesis is
  obtained by performing linear regression using the two data sets
  that respectively give the minimal and maximal slopes among all
  realizations if fitted individually.}
\begin{ruledtabular}
\begin{tabular}{lcccccc}
\textrm{$\phi_{2D}$} & \textrm{$N_{eject}^{tot}$} &
\textrm{$N_{push}^{tot}$} & \textrm{$\Delta L^{max}$} &
\textrm{$\Delta N_{push}^{max}$} & \textrm{$m_{\Delta L}$} &
\textrm{$m_{\Delta N}$} \\ \colrule (simulation) & & & & & &
\\ $0.762$ & $50$ & $15674.2 \pm 5450.0$ & $0.10 \pm 0.01 D_d$ &
$2668.7 \pm 2058.1$ & $-0.81 \pm 0.19$ ($-0.77$) & $-0.86 \pm 0.11$
($-0.88$) \\ $0.762$ w/ PBC & $23.7 \pm 3.0$ & $7796.7 \pm 1152.8$ &
$0.17 \pm 0.04 D_d$ & $1341.1 \pm 701.2$ & $-0.85 \pm 0.16$ ($-0.78$)
& $-0.73 \pm 0.20$ ($-0.93$) \\ $0.686$ & $26.4 \pm 2.8$ & $8693.8 \pm
869.4$ & $0.25 \pm 0.03 D_d$ & $1261.5 \pm 438.6$ & $-0.87 \pm 0.17$
($-0.73$) & $-0.75 \pm 0.10$ ($-0.81$) \\ $0.610$ & $20.1 \pm 2.5$ &
$4446.8 \pm 802.2$ & $0.24 \pm 0.02 D_d$ & $754.4 \pm 249.5$ & $-0.88
\pm 0.09$ ($-0.91$) & $-0.79 \pm 0.09$ ($-0.84$) \\ \colrule
(experiment) & & & & & & \\ $0.762$ & $50$ & $105333.3 \pm 39570.1$ &
$0.21 \pm 0.02 D_d$ & $14351.9 \pm 4100.7$ & $-0.59 \pm 0.13$
($-0.67$) & $-1.17 \pm 0.06$ ($-1.15$) \\ $0.686$ & $49 \pm 1.4$ &
$55670.4 \pm 14820.1$ & $0.33 \pm 0.01 D_d$ & $10955.6 \pm 1726.0$ &
$-0.69 \pm 0.08$ ($-0.64$) & $-1.20 \pm 0.18$ ($-1.16$) \\ $0.610$ &
$35 \pm 2.2$ & $19314.8 \pm 7731.3$ & $0.30 \pm 0.06 D_d$ & $4707.4
\pm 1960.4$ & $-1.02 \pm 0.46$ ($-0.86$) & $-1.40 \pm 0.24$ ($-1.27$)
\\
\end{tabular}
\end{ruledtabular}
\end{table*}

\begin{figure}
\includegraphics[width=0.39\textwidth]{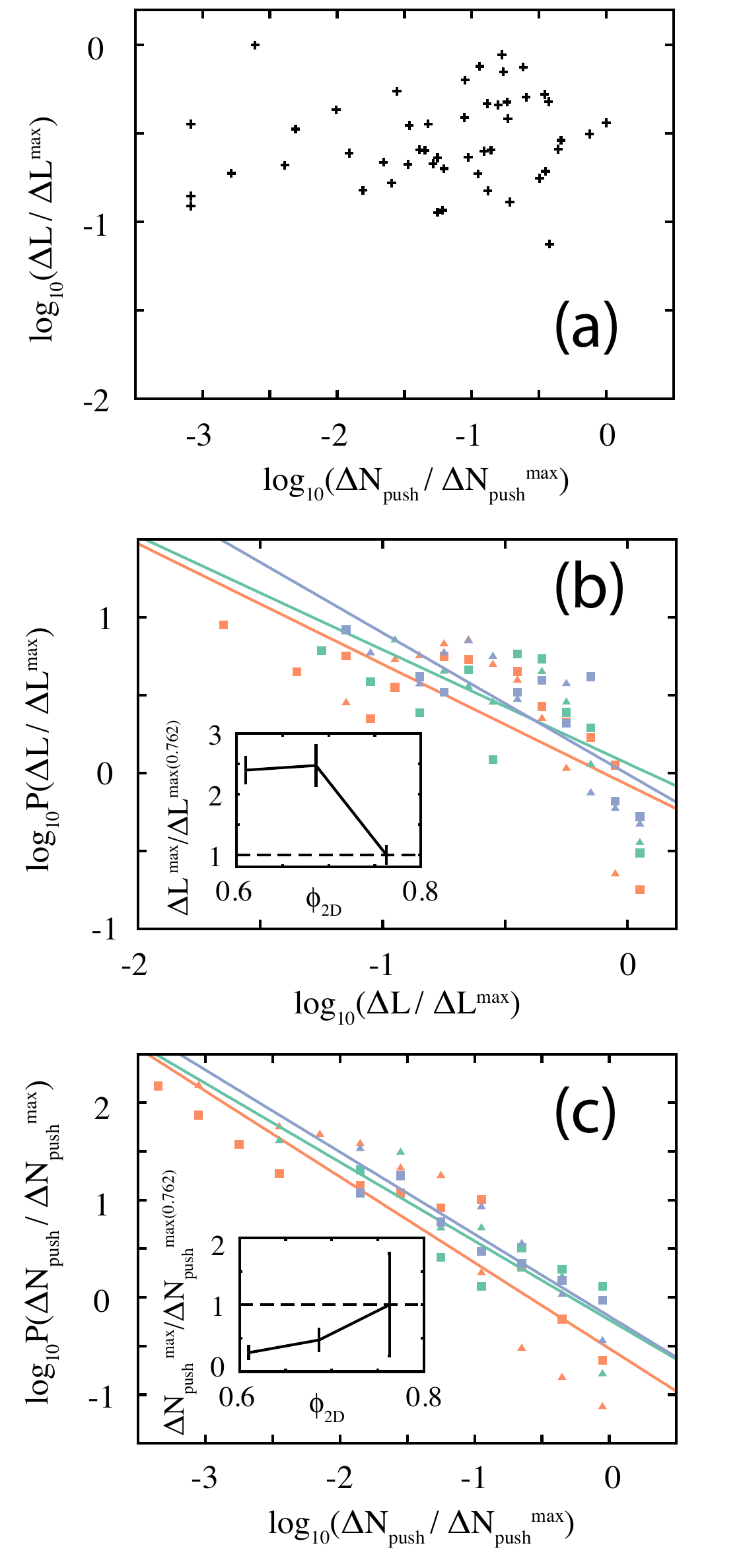}
\caption{\label{fig:Simu_N50moved_dist_dur_corr} (a) The normalized
  moving distance $\Delta L / \Delta L^{max}$ between two consecutive
  ejections, plotted against the associated normalized time interval
  $\Delta N_{push} / N_{push}^{max}$. System packed with $\phi_{2D}
  \approx 0.762$ with side walls. (b) The probability density
  functions $P$ of $\Delta L / \Delta L^{max}$ with $\phi_{2D} \approx
  0.762$ (orange), $0.686$ (green), and $0.610$ (blue) and the
  corresponding linear regression lines of slope $m_{\Delta L}$. (c)
  The probability density functions $P$ of $\Delta N_{push} / \Delta
  N_{push}^{max}$ and the corresponding linear regression lines of
  slope $m_{\Delta N}$. A linear regression line shown in (b) and (c)
  is obtained using the two data sets that respectively give the
  minimal (square) and maximal (triangle) slopes among 10 realizations
  if fitted individually. The insets show averaged $\Delta L^{max}$
  and $\Delta N_{push}^{max}$, respectively, normalized by their
  corresponding values at $\phi_{2D} \approx 0.762$. Each mean and its
  variation in the insets are obtained using $10$ realizations. The
  values of $\Delta L^{max}$, $\Delta N_{push}^{max}$, $m_{\Delta L}$
  and $m_{\Delta N}$ are listed in Table \ref{tab:sim_exp_results}.}
\end{figure}

The distributions of the moving distance $\Delta L$ and the associated
time interval $\Delta N_{push}$, after being normalized by the maximum
value, between consecutive ejections with $\phi_{2D} \approx 0.762$ is
further examined in Fig. \ref{fig:Simu_N50moved_dist_dur_corr}(a),
which shows a weak correlation between the two quantities. The
probability distributions of normalized $\Delta L$ and $\Delta
N_{push}$ between consecutive ejections with $\phi_{2D} \approx
0.762$, $0.686$, and $0.610$ reveal a power-law behavior, as shown in
Fig. \ref{fig:Simu_N50moved_dist_dur_corr}(b) and
Fig. \ref{fig:Simu_N50moved_dist_dur_corr}(c), respectively. How the
data are extracted and processed to characterize the power-law
behavior are detailed in the caption of Table
\ref{tab:sim_exp_results}. The fitted values suggest that both the
spans of $\Delta L$ and $\Delta N_{push}$ over which the power laws
hold increase with increasing $\phi_{2D}$, while $\Delta L^{max}$
decreases and $\Delta N_{push}^{max}$ increases, respectively.  The
power laws with global exponents $m_{\Delta L}$ and $m_{\Delta N}$
around or close to one in both space and time suggest that digging
within a densely-packed granular environment may be a critical
phenomenon (besides, we notice that the probability distribution of
normalized $\Delta L$ exhibits a transition between two local slopes
around $\Delta L \approx \Delta x$, where $\Delta x \approx {\alpha
  _x}\sigma = 0.025D_d$ is the size of the small random displacement
defined in Eqn. \ref{digger_bump}. We will investigate this transition
using smaller values of $\Delta x$ in our future work).

\subsubsection{The numerical digger in a layer of cylindrical particles with the periodic boundary conditions}
\label{DEM_cylindrical_part_w_pbc}

\begin{figure}
\includegraphics[width=0.34\textwidth]{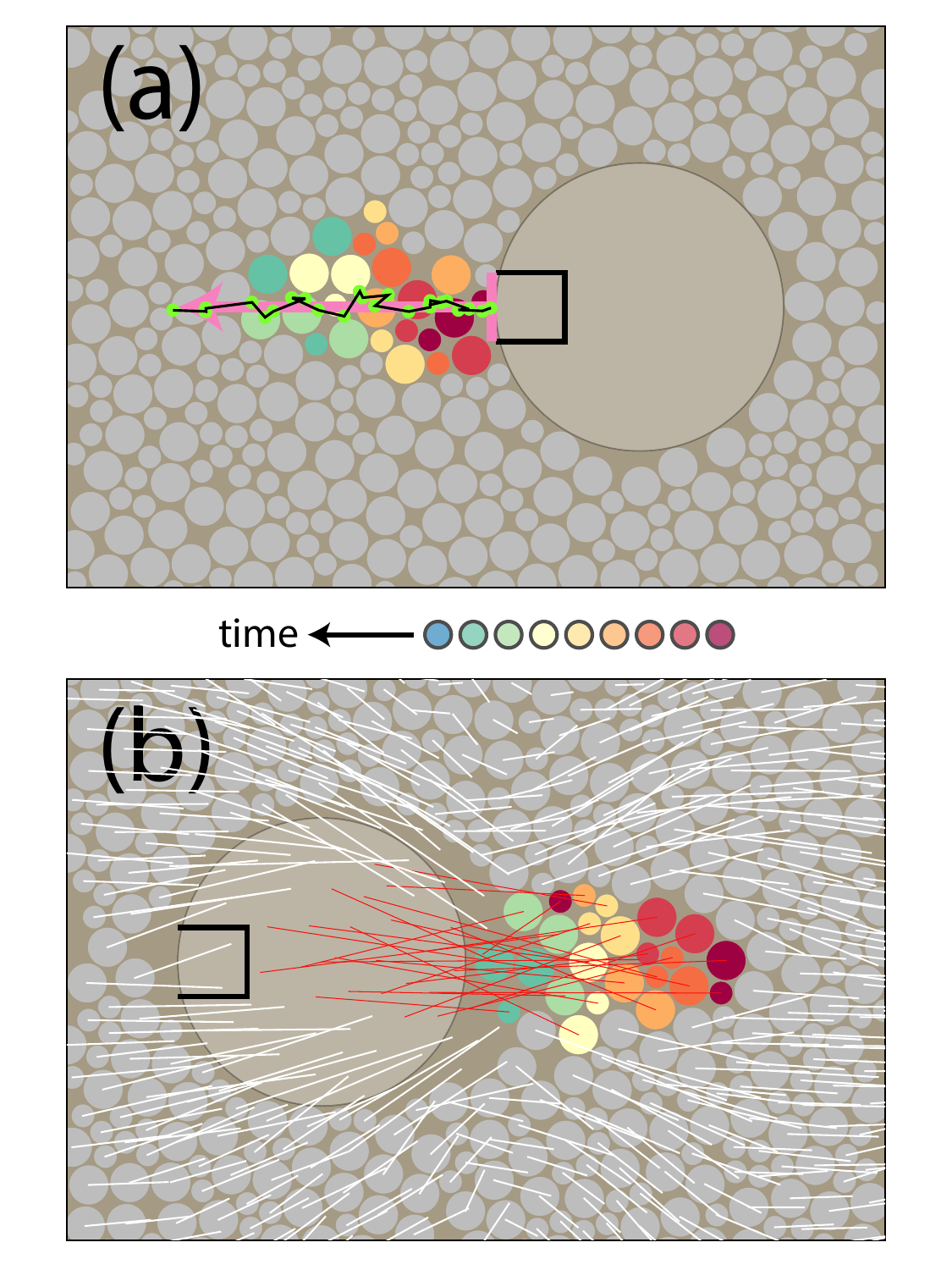}
\caption{\label{fig:Simu_NXXmoved_w_pbc} The same as
  Fig. \ref{fig:Simu_N50moved}(a1) and (b1), except the periodic
  boundary conditions in both $x$ and $y$ directions are implemented
  with $\phi_{2D} \approx 0.762$.}
\end{figure}

Ideally, a power-law behavior and the related critical phenomenon
should be tested in the thermodynamic limit. However, in the
experimental setup, the digger diameter has $D_d = 21\ \mathrm{cm}$,
while the container measures only $37\ \mathrm{cm}$ ($W_c$) by
$54\ \mathrm{cm}$ ($L_c$). Finite-size and boundary effects on our
results are inevitable, for force networks and particle rearrangements
in dense packings are known to be strongly wall-dominated at this
relative size scale. We tackle this issue by implementing periodic
boundary conditions (PBC) in both $x$ and $y$ directions for the
granular packing prepared in
Sec. \ref{DEM_cylindrical_part_wo_pbc}. The system configurations with
$\phi_{2D} \approx 0.762$ before and after the digger has moved a
distance comparable to its size are shown in
Fig. \ref{fig:Simu_NXXmoved_w_pbc}(a) and (b), respectively. Without
the side walls of the container, the number of particles that have to
be relocated is reduced by half. However, due to the nature frequent
particle-digger interactions in dense granular packings, the PBC
digger cannot move extremely further than the confined digger without
PBC.

\begin{figure}
\includegraphics[width=0.39\textwidth]{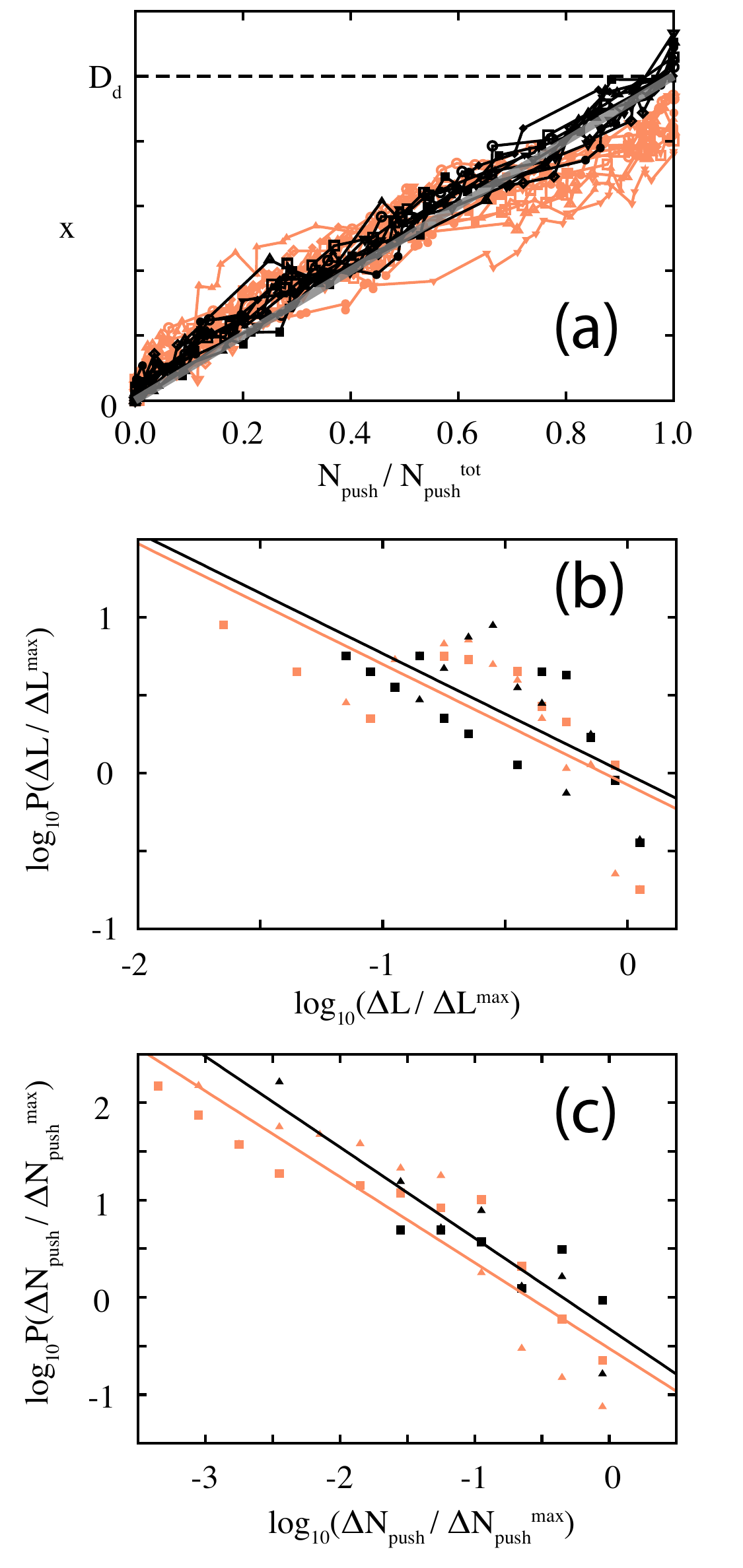}
\caption{\label{fig:Simu_NXXmoved_dist_dur_w_pbc} Implementing the
  periodic boundary conditions in both $x$ and $y$ directions for a
  system of $\phi_{2D} \approx 0.762$ to study (a) digger dynamics
  using $10$ realizations (black), compared to those with side walls
  (orange), and its statistical behaviors in view of (b) advancement
  and (c) number of push attempts between consecutive ejections,
  respectively. The values of $\Delta L^{max}$, $\Delta
  N_{push}^{max}$, and the linear regression line slopes $m_{\Delta
    L}$ and $m_{\Delta N}$ are listed in Table
  \ref{tab:sim_exp_results}.}
\end{figure}

In Fig. \ref{fig:Simu_NXXmoved_dist_dur_w_pbc}(a), we plot the
trajectories of the digger along the global reference direction $x$
against the normalized $N_{push}$. By comparing all the $10$ results
without (orange lines) and with (black lines) periodic boundary
conditions, it is clear that the numerical digger confined by the
container walls slows down after moving a distance of about $0.6D_d$
closing up to the downstream wall. Besides, after relocating $50$
particles, the longest distance that the confined digger can move is
about $0.9D_d$ instead of $D_d$ of the digger subject to no walls,
which uses roughly only half of the $N_{push}^{tot}$ needed in the
confined case (see Table \ref{tab:sim_exp_results}). In
Fig. \ref{fig:Simu_NXXmoved_dist_dur_w_pbc}(b) and (c), we show the
distributions of normalized $\Delta L$ and $\Delta N_{push}$ between
consecutive ejections, which reveal the familiar power-law behavior as
those observed from the experiments and the simulations with side
walls. Removing the container walls shrinks the spans of $\Delta L$
and $\Delta N_{push}$ over which the power laws hold but does not
destroy them, which indicates that the observed power-law behavior and
the related critical phenomenon is robust in the thermodynamic limit
and results primarily from the interactions between the digger and the
densely packed particles.

The potential critical phenomenon can be perceived as follows. The
force chain network that can support applied loads around the digger
changes during the digging process. Presumably, the most pronounced
changes of the force chain network occur in where the free space is
the largest and the local packing is the most fragile. Therefore
looser granular obstacles near the front recess of the digger are more
likely to be captured and relocated to the rear of the digger. As the
digging continues, there would be a series of reconstructions or even
avalanches of load-bearing force chains \cite{claudin98} and is
reflected in the power-law distribution of moving distances and time
intervals between consecutive ejections by the digger. In the next
section, we verify the power-law behavior experimentally using an
automated digger.

\subsection{The automated digger in a layer of cylindrical particles}
\label{automated_digger}

\begin{figure}
\includegraphics[width=0.39\textwidth]{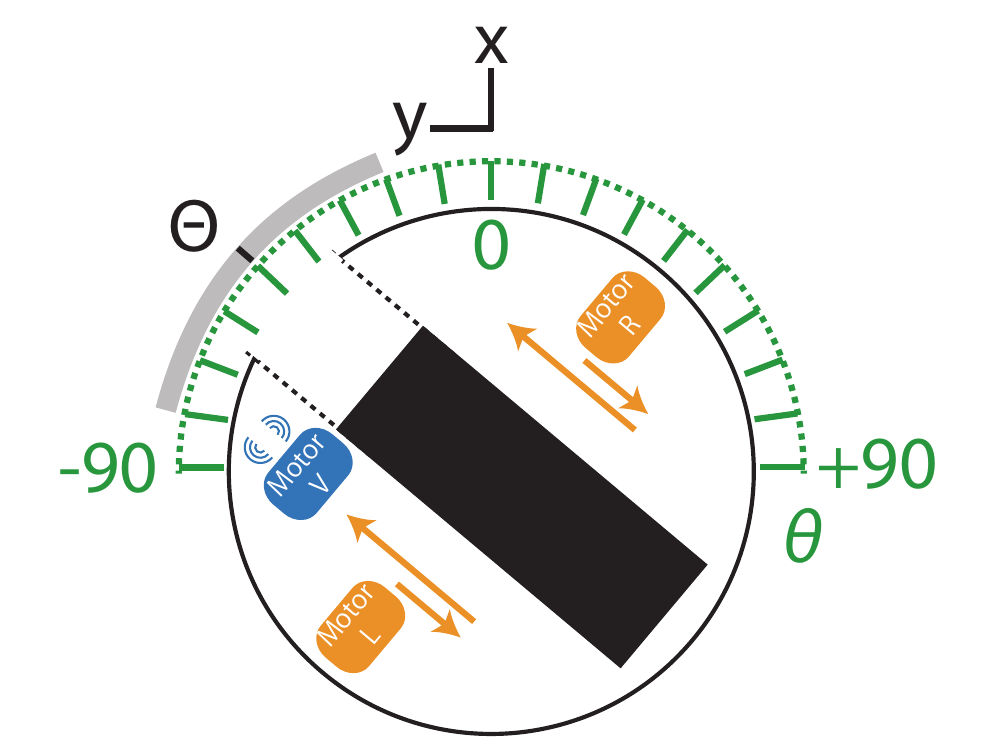}
\caption{\label{fig:microbit_protractor} Schematic of the automated
  digger. The heading angle of the digger $\Theta$ (black) is defined
  on the angular coordinate $\theta$ (green), ranging anywhere from
  $-90$ degrees to $90$ degrees with $0$ degrees being fixed along the
  global reference direction $x$. The figure shows a representative
  digging range of $\Theta$ with $\theta_1=-50$ degrees combined with
  multiple choices of $\theta_2$ (spanned by the gray arc). For a
  chosen $\Theta$, the digger repetitively pushes forward and retreats
  (orange arrows) over a total attempt time of one minute. The digger
  keeps trying different values of $\Theta$ until it captures at least
  one cylinder into its front recess naturally, which is the automated
  version of step 1 shown in Fig. \ref{fig:schematic}(b1). The rest
  steps of the moving procedure remain unchanged.}
\end{figure}

To examine whether we can experimentally reproduce the power laws
observed in simulation, we automate the manual digger by periodically
randomizing its digging direction. This removes human judgement from
the selection of which granular obstacles to relocate. For automated
control, an angle coordinate system $\theta$ is defined along the
$x$-axis of the global coordinate system as marked in
Fig. \ref{fig:setup} and Fig. \ref{fig:microbit_protractor}. To ensure
that the automated digger does not get stuck in one place, we assign a
much wider range of heading angle of pushing, $\Theta = -90$ to $90$
degrees, than that used for the manual operation. This also allows the
digger to achieve perceivable moving distances within reasonable
experimental time frames. The automated digger proceeds by a heading
angle $\Theta = \theta_1 + \theta_2$, chosen randomly among $-90$,
$-80$, ..., $0$, ..., or $90$ for $\theta_1$ and $-30$, $-20$, ...,
$0$, ..., or $30$ for $\theta_2$, both set with a constant $10$
degrees increment. We discard values of $\Theta$ smaller than $-90$
and greater than $90$ so that the digger does not attempt to push
backward. At each $\Theta$, within an overall duration of one minute,
the digger repetitively pushes forward for $\Delta t_f=60$
milliseconds (equivalent to a nominal moving distance $\Delta x_f
\approx 0.045D_d$) and then retreats (moves backward) for $\Delta
t_b=30$ milliseconds with concurrent vibrations at a frequency $f
\approx 200$ Hz. The low-amplitude vibrations are used in the
automated experiments to reduce local interparticle friction around
the recess of the digger, so that our experiments can be completed
within a reasonable laboratory time. The repetitive back and forth
movement of the digger when it makes a quasi-static push attempt,
controlled by a micro:bit microcontroller, is to mimic the behavior of
the manually-operated digger. For a given $\theta_1$, the digger
repeats the process seven times with randomly chosen $\theta_2$ so
that in principle it can explore the neighborhood of $\theta_1$ for as
long as seven minutes. The digger keeps changing the value of $\Theta$
until it captures at least one cylinder, as shown by the movie clip
provided in the Supplemental Material \cite{supp4}. The process of
particle ejection remains unchanged and after an ejection the whole
automated process starts over again.

In comparison to the numerical digger, the automated digger randomizes
its digging direction $\Theta$ and makes quasi-static push attempts,
by a microcomputer, which is similar to the random displacements of
the numerical digger. Besides, even though we automated the digger by
periodically assigning a random digging direction to it, which
involves rotations, the average heading angle of the automated digger
is still at $\Theta = 0$ degrees, which is consistent with the
numerical algorithm.

\begin{figure}
\includegraphics[width=0.30\textwidth]{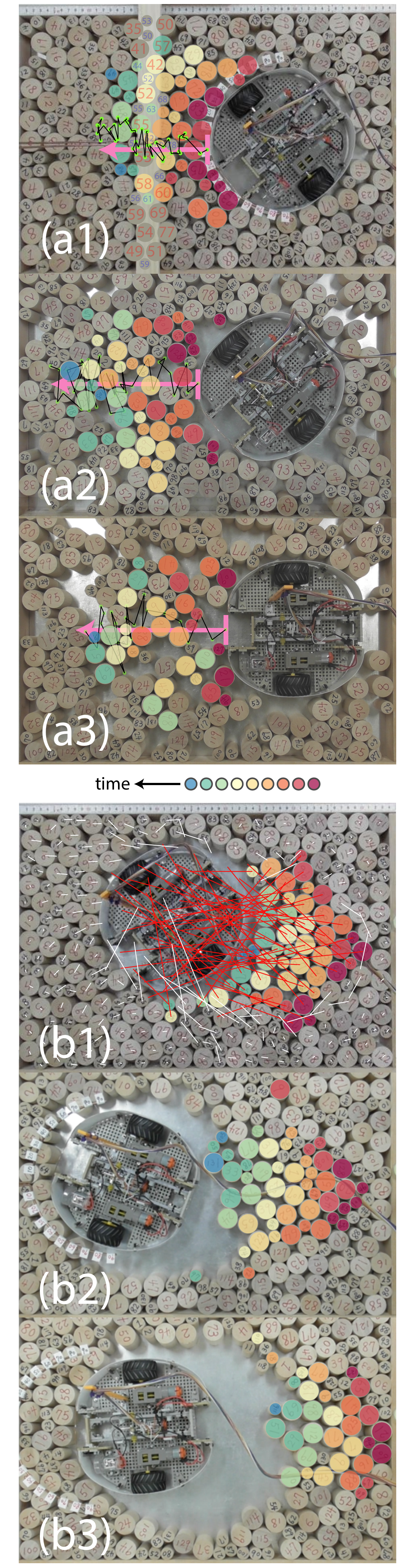}
\caption{\label{fig:Exp_cylinder_Microbit_N50moved} Same as
  Fig. \ref{fig:Simu_N50moved}, except the digger is automated. System
  packed with $\phi_{2D} \approx 0.762$ (a1, b1), $0.686$ (a2, b2),
  and $0.610$ (a3, b3).}
\end{figure}

The system configurations with $\phi_{2D} \approx 0.762$, $0.686$, and
$0.610$ before and after the automated digger has moved a distance
comparable to its size are shown in
Fig. \ref{fig:Exp_cylinder_Microbit_N50moved}(a) and
Fig. \ref{fig:Exp_cylinder_Microbit_N50moved}(b), respectively. Like
the results of the numerical autonomous digger in
Fig. \ref{fig:Simu_N50moved}, reducing $\phi_{2D}$ also increases the
void size in the particle bed behind the automated digger and
therefore the digger's dynamics and statistical behaviors are
similar. The wider heading angle ($\Theta = -90$ to $90$ degrees)
generally results in a wider spatial distribution of relocated
particles and a more scattered trajectory than those of the numerical
digger, whose heading angle is limited to around $\Theta = 0$ degrees.

\begin{figure}
\includegraphics[width=0.39\textwidth]{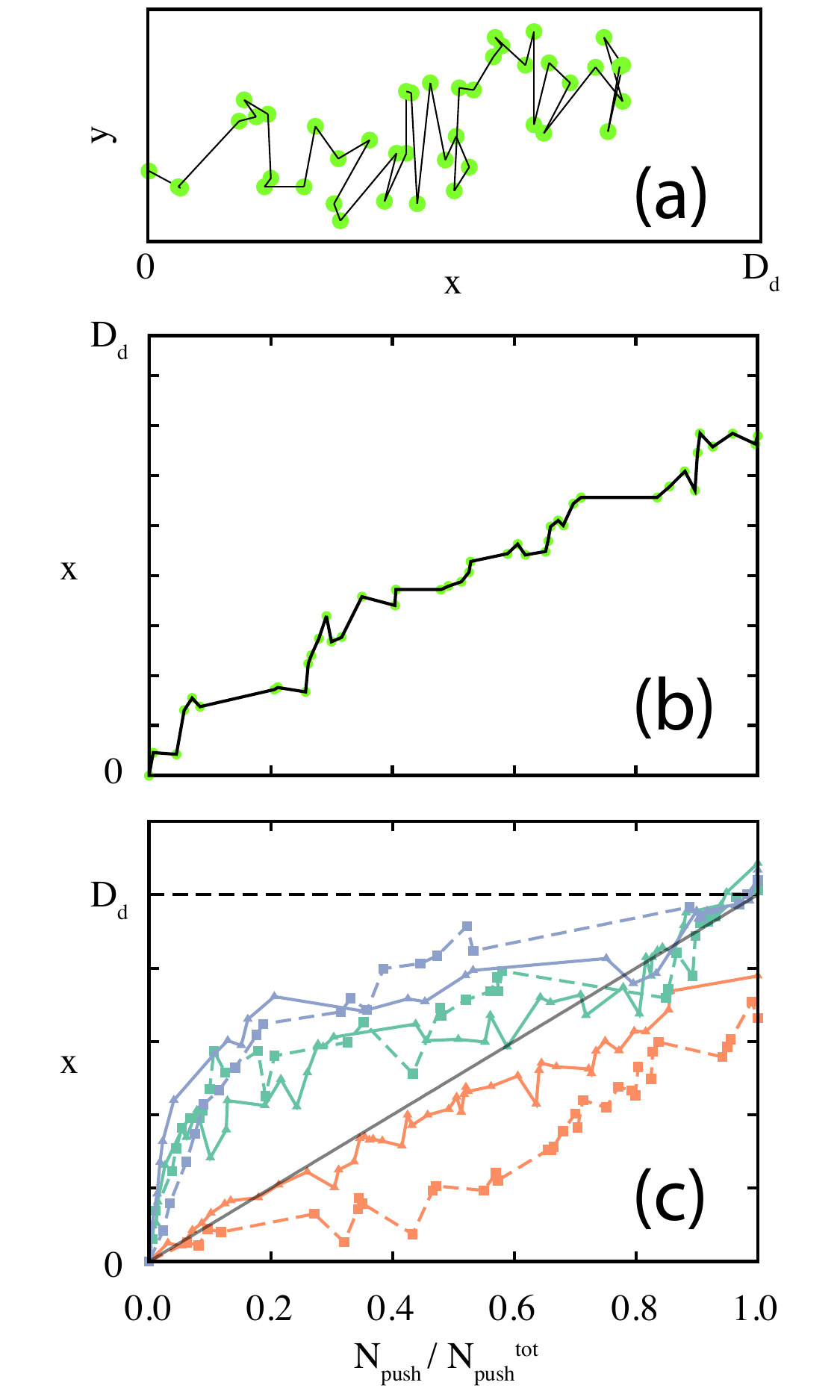}
\caption{\label{fig:Exp_Microbit_N50moved_caging} Same as
  Fig. \ref{fig:Simu_N50moved_caging}, except the digger is automated
  and only the trajectories of the digger after each ejection are
  shown. $N_{push}^{tot} = T_{push}^{tot}/(\Delta t_f + \Delta t_b)$,
  where $ \Delta t_f + \Delta t_b = 90 \text{ms}$, are listed in Table
  \ref{tab:sim_exp_results}.}
\end{figure}

We quantify the automated digger's mobility by studying the spatial
trajectory of the digger after each ejection. Specifically, we define
the location where the digger starts to rotate to eject the captured
granular obstacles as the final (relaxed) position of an
ejection. Even though this definition is slightly different from the
one based on a well-defined energy minimization criterion used in DEM
simulations, it eliminates the potential human influence of deciding
on where to eject the captured particles introduced through the rest
of the moving procedure steps. The trajectory with $\phi_{2D} \approx
0.762$ is shown in Fig. \ref{fig:Exp_Microbit_N50moved_caging}(a). We
also plot the position of the digger along the global reference
direction $x$ against the total number of pushes $N_{push} \equiv
T_{push}/(\Delta t_f + \Delta t_b)$, where $T_{push}$ is the total
pushing time by the digger. As with our numerical analysis, $T_{push}$
only includes durations of capturing granular obstacles. The temporal
evolution of the digger’s advancement with $\phi_{2D} \approx 0.762$
in the global reference direction $x$, shown in
Fig. \ref{fig:Exp_Microbit_N50moved_caging}(b), presents a similar
stepwise curve to the simulated trajectory shown in
Fig. \ref{fig:Simu_N50moved_caging}(b) and confirms the effectiveness
of the automated digging mechanism. In
Fig. \ref{fig:Exp_Microbit_N50moved_caging}(c), we plot the
advancement of the digger along the $x$ direction against $N_{push}$
with $\phi_{2D} \approx 0.762$, $0.686$, and $0.610$. Since the
attempted $N_{push}^{tot}$ varies significantly from the experiments
and simulations, we normalize each physical quantity by its own
corresponding maximum, and the overall trend is similar to the one
found numerically in Fig. \ref{fig:Simu_N50moved_caging}(c). This
resemblance should suggest that our analyses are not influenced by the
difference between the numerical and automated diggers, and the
finding is generic.

\begin{figure}
\includegraphics[width=0.39\textwidth]{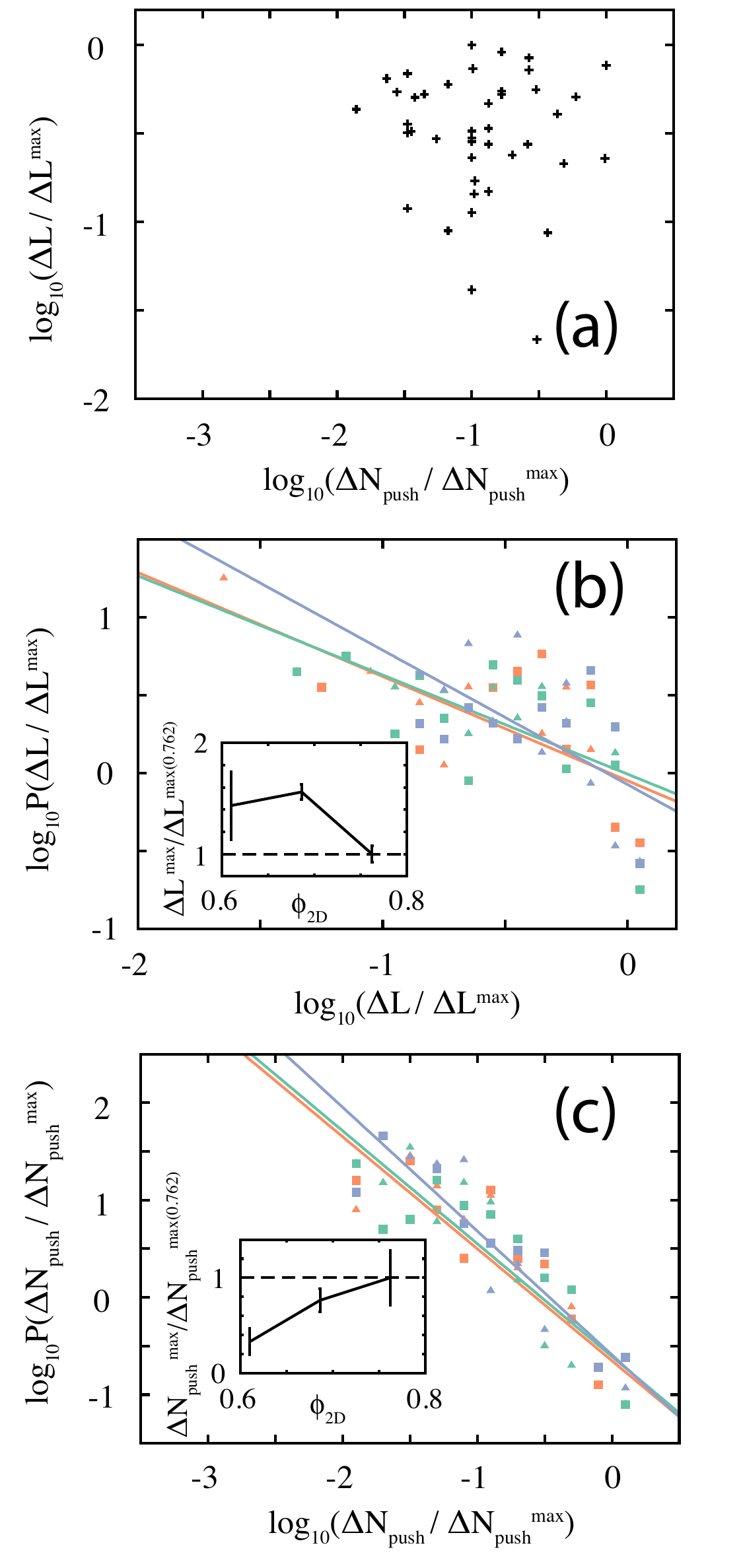}
\caption{\label{fig:Exp_Microbit_N50moved_dist_dur_corr} Same as
  Fig. \ref{fig:Simu_N50moved_dist_dur_corr}, except the digger is
  automated with $\phi_{2D} \approx 0.762$ (orange), $0.686$ (green),
  and $0.610$ (blue). Each mean and its variation in the insets are
  obtained using $3$ realizations. The values of $\Delta L^{max}$,
  $\Delta N_{push}^{max}$, and the linear regression line slopes
  $m_{\Delta L}$ and $m_{\Delta N}$ are listed in Table
  \ref{tab:sim_exp_results}.}
\end{figure}

We also perform spatial and temporal analyses similar to that on the
numerical results in Fig. \ref{fig:Simu_N50moved_dist_dur_corr}. The
moving distance $\Delta L$ between consecutive cylinder ejections with
$\phi_{2D} \approx 0.762$ is plotted against the associated time
interval $\Delta N_{push}$ in
Fig. \ref{fig:Exp_Microbit_N50moved_dist_dur_corr}(a) after
normalization with respective maximum values. The distributions of
normalized $\Delta L$ with $\phi_{2D} \approx 0.762$, $0.686$, and
$0.610$ are shown in
Fig. \ref{fig:Exp_Microbit_N50moved_dist_dur_corr}(b). We observe a
similar power-law behavior like the one shown in
Fig. \ref{fig:Simu_N50moved_dist_dur_corr}(b), except across a
narrower range of $\Delta L$. The probability distribution of
normalized $\Delta L$ exhibits a transition between two local slopes
when $\Delta L$ is on the order of $\Delta x_f$.  In
Fig. \ref{fig:Exp_Microbit_N50moved_dist_dur_corr}(c), we plot the
distributions of normalized $\Delta N_{push}$ and again observe a
power-law behavior, like the one shown in
Fig. \ref{fig:Simu_N50moved_dist_dur_corr}(c), except also across a
narrower range of $\Delta N_{push}$. The spatial and temporal power
laws observed for the two digging dynamical variables, $\Delta L$ and
$\Delta N_{push}$, should suffice to confirm the digging mechanism is
robust and the numerical digger exhibits similar dynamic behavior to
that of the mechanical one.

\section{Conclusions}
\label{conclusions}

In this study, we propose a digger capable of moving within a layer of
densely-packed coarse granular particles with an average grain size
about one tenth as large as the digger. This is intrinsically
different from previous bio-inspired autonomous diggers mimicking the
intricate movements of organisms living in sand, and to move forward,
they rely on the yield and the fluidization of surrounding grains that
are much smaller than the digger.  Our digger has a circular shape
with a recess in the front and an optional vibrating function that
allows it to capture and then eject sizable granular particles. The
geometric design of the digger in principle can be extended to a 3D
environment, where the digger may have a spherical shape. This design
empowers our digger to navigate working environments that are hard to
fluidize globally due to comparably strong interparticle friction and
hence difficult for bio-inspired diggers to negotiate.

We first manually tested our digger within simple-shaped wood
cylinders. We then changed the granular medium to irregular-shaped
gravels to test the influence of particle shape on the mobility
performance of the digger. The complicated shape of gravel requires
the digger to activate its vibrating function to loosen interlocked
granular particles. While the digger moves much slower in
irregular-shaped gravels than in regular-shaped cylinders, the
distribution of relocated particles before and after the digger's
advancement to a distance comparable to its size stays almost the
same. This similarity proves the feasibility of the four-step
relocation-for-advancement mechanism uniquely achieved by the current
design.

After finishing the manual tests, we compared our experimental results
with the DEM simulation that is independent of the human judgement on
the operation of the digger. The strong similarity between the
experimental and numerical results indicates that the digger with its
mobility mechanism can potentially become fully autonomous.  However,
we have also found that an operator's dexterity and flexibility can
help the digger move more efficiently by adapting to the ever-changing
on-site situation. To evaluate the digger's mobility and efficiency,
we analyzed moving distances and time intervals between consecutive
ejections. Our analyses show that the distribution of both quantities
follow a power-law. Reducing $\phi$ shrinks the spatial and temporal
spans over which the power laws hold. To confirm this finding
experimentally with minimal human influence, we automated the digger
by periodically assigning a random digging direction to
it. Strikingly, the digging strategy succeeded in reality, and we also
observe similar spatio-temporal power laws. The power laws with
exponents around or close to one found both numerically and
experimentally suggest that digging within a densely-packed coarse
granular environment could be a critical phenomenon.

In this study, the vibrating function operates at a fixed frequency
about $200$ Hz. However, in general, the vibrating frequency could be
optimized, as indicated by studies of vibrating machines such as
ballast tampers which use claw-like tines vibrating at an optimal
frequency of $35$ Hz to fluidize the ballast for railroad track
maintenance. Too low a frequency is ineffective; too high one
increases the impact force acting on the vibrated granular particles
and may cause them to break or further interlocked \cite{shi20,
  guo21}. Besides, we could also alternate the stiffness of the
circular periphery of the digger to facilitate its movement. For
example, by armoring the periphery of the digger with soft inflatable
objects, whose stiffness can become stiffer or softer during inflation
due to strain softening \cite{pashine25}.  We will explore the optimal
vibrating frequency and the periphery stiffness of the digger in our
future work.

\begin{figure}
\includegraphics[width=0.45\textwidth]{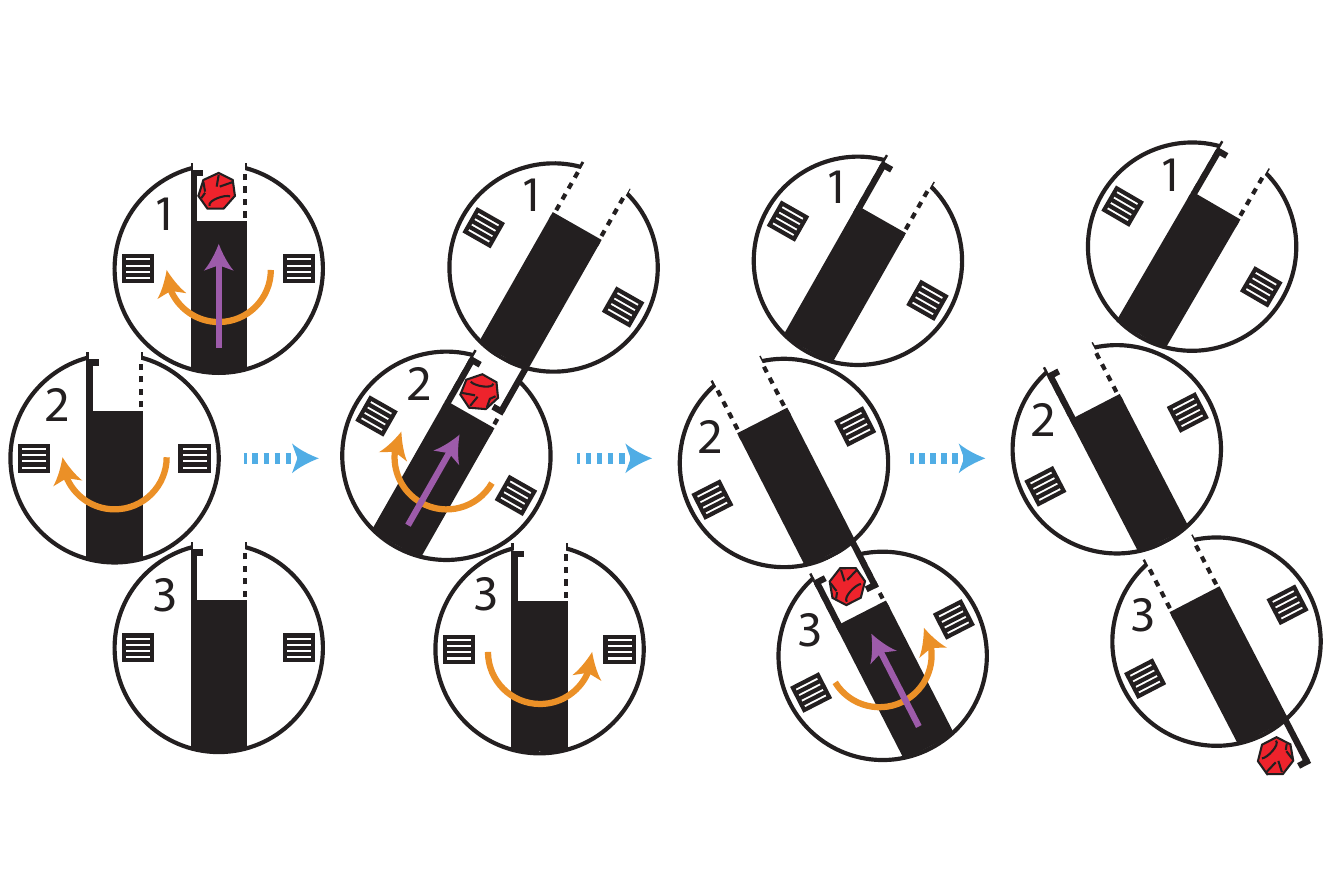}
\caption{\label{fig:fireman} Transporting granular particles through
  collaboration of multiple diggers.}
\end{figure}

Potential applications of the digger include rescue missions in debris
after an earthquake or a mining accident.  For example, as shown in
Fig. \ref{fig:fireman}, we can use multiple diggers for transporting
debris. An ship version of the digger surrounded by floating obstacles
such as drift ice can be used for ice management around offshore
structures in frozen seas. We hope to learn more about this unexplored
densely-packed granular system with small ratio of digger size to
characteristic grain size in time to come.

\section{acknowledgments}
G.J.G. acknowledges financial support from Shizuoka University and
Hamamatsu Foundation for Science and Technology Promotion
(Japan). F.-L.Y. acknowledges Grants MOST-113-2923-E-002-009 and
MOST-112-2221-E-002-137-MY3 from Ministry of Science and Technology
(Taiwan).

\bibliography{paper}

\end{document}